\documentclass[aps,pra,floatfix,amsmath,amssymb,superscriptaddress,showpacs,showkeys,onecolumn,10pt]{revtex4-2}
\usepackage[caption=false]{subfig}
\usepackage{graphicx,bm,color}
\usepackage{amsfonts}
\usepackage{color}
\usepackage{multirow}
\usepackage{lipsum, color}
\usepackage{mathtools}
\DeclarePairedDelimiter{\ceil}{\lceil}{\rceil}
\usepackage{amsthm}

\newtheorem{definition}{Definition}
\newtheorem{Theorem}{Theorem}
\newtheorem{Lemma}{Lemma}
\usepackage{braket}
\usepackage{hyperref}
\hypersetup{
    colorlinks=true,
    linkcolor=blue,
    filecolor=blue,      
    urlcolor=blue,
}
\usepackage{graphicx}
\begin{document}

\frenchspacing
\title{Generalized Teleportation Fidelity and  Singlet Fraction and their Relation for  (In)-distinguishable Particles and Its Applications}
\author{Soumya Das}
\email{s.das2@tue.nl}
 \affiliation{Department of Mathematics and Computer Science, Eindhoven University of Technology}
\author{Goutam Paul}%
 \email{goutam.paul@isical.ac.in}
\affiliation{ 
Cryptology and Security Research Unit, Indian Statistical Institute, Kolkata 700108, India
}%
\author{Anindya Banerji}
\email{cqtab@nus.edu.sg}
 \affiliation{Centre for Quantum Technologies, National University of Singapore}
\begin{abstract}
Quantum teleportation efficiently transfers quantum information between distant locations by utilizing a pre-established composite system.
Assessing the effectiveness of teleportation hinges on its fidelity, representing the similarity between input and output states. This fidelity, in turn, relies on a singlet fraction, quantifying the resemblance of the composite system to maximally entangled states. 
The relation between teleportation fidelity and singlet fraction given by \href{https://doi.org/10.1103/PhysRevA.60.1888}{[Horodecki \textit{et al}., Phy. Rev. A \textbf{60}, 1888 (1999)]} does not hold for distinguishable particles with multiple degrees of freedom or indistinguishable particles with single or multiple degrees of freedom. In this paper, we propose generalized expressions for teleportation fidelity and singlet fraction and derive their relations, applicable for both distinguishable and indistinguishable particles with single or multiple degrees of freedom. We derive an upper bound for the generalized singlet fraction for distinguishable particles using the monogamy of singlet fraction by \href{https://doi.org/10.1103/PhysRevLett.103.050501}{[Kay \textit{et al.} Phys. Rev. Lett. \textbf{103}, 050501 (2009)]}.  We also show how our relation helps to characterize different types of composite states in terms of their distinguishability, separability, presence of maximally entangled structure, and the number of degrees of freedom. We complement our theory with two practical illustrations. First, we demonstrate two counter-intuitive values of generalized singlet fraction using our optical circuit and the circuit of \href{https://doi.org/10.1103/PhysRevLett.120.050404}{[Li \textit{et al.}, Phys. Rev. Lett. \textbf{120}, 050404 (2018)]}. Finally, we show that using an additional degree of freedom as an ancilla instead of a particle can be advantageous in quantum cryptographic protocols. 
\end{abstract}


\maketitle

\maketitle

\section{\label{sec:level1} INTRODUCTION }
In recent years, quantum information processing 
with independently prepared identical particles, featuring multiple degrees of freedom, has emerged as a promising research frontier. However, when these particles become indistinguishable due to the spatial overlap of their wave functions, the fundamental principles and applications of quantum information processing using distinguishable particles no longer hold For example, it was known that quantum teleportation could be performed with unit fidelity using distinguishable particles~\cite{QT93}. But, Das \textit{et al.}~\cite{Das20} have proved by a \textit{no-go} theorem that unit fidelity quantum teleportation using indistinguishable particles will violate the no-signaling principle. Entanglement of indistinguishable particles finds applications in Bose-Einstein condensate~\cite{Morsch06,Esteve08}, quantum metrology~\cite{Giovannetti06,Benatti11}, quantum dots~\cite{Petta05,Tan15}, ultracold atomic gases~\cite{Leibfried03}, etc.

Quantum entanglement is generally encoded in particles' degrees of freedom (DoFs) like spin, angular momentum, etc. Initially, the proposed applications of quantum information utilized a single DOF of the particles involved. However, with the advancement of technology,  multiple DoFs of a single particle can be addressed reliably.  Thus researchers are moving from single-DoF multi-particle scenarios to multi-DoF multi-particle scenarios~\cite{Zukowski91,Kwait97,Ma09,Nagali09,Jeong14,Andersen15,Zhang16,Camalet17,Camalet18,HHNL} that find applications in efficient Bell measurement~\cite{Schuck06}, superdense coding~\cite{Graham15}, state
purification~\cite{Simon02},  quantum computing~\cite{Chen07}, quantum  networks~\cite{Nagali10}, etc. Further, when particles become indistinguishable, some properties or applications may be different from the distinguishable ones. For example, Das \textit{et al.}~\cite{Das20} has shown that hyper-hybrid entanglement, a special case of multi-DoF entanglement can only be produced using indistinguishable particles but cannot exist for distinguishable particles. 

Quantum teleportation fidelity~\cite{Jozsa94} plays an important role in quantum information for distinguishable and indistinguishable particles. It measures the closeness between the state  to be teleported and the output state  obtained after teleportation.      
\begin{definition}
    The teleportation fidelity $(f)$ for the state $\rho^{in}$ to be teleported and the output state $\rho^{out}$ obtained after teleportation is defined by
\begin{equation} \label{Tf}
 f:= \text{Tr}\sqrt{\sqrt{\rho^{in}}\rho^{out}\sqrt{\rho^{in}}}.
\end{equation} 
\end{definition}
It is related to the teleportation channel $\rho$ whose quality is measured by the singlet fraction~\cite{Horodecki99}, i.e., the maximum overlap of $\rho$ with maximally entangled states, defined as 
\begin{definition}
    The singlet fraction $(F)$  with respect to any state $\rho$ is defined by
\begin{equation} \label{SFold}
F:=\max_{\psi} \braket{\psi \mid \rho \mid \psi},
\end{equation}
where $\ket{\psi}$ varies over all maximally entangled states.  
\end{definition}
The relation between the above two quantities for any bipartite quantum state acting on $\mathbb{C}^d \otimes \mathbb{C}^d$ under local operations and classical communications is	 	
derived in~\cite{Horodecki99} as
\begin{equation} \label{f_F_par}
f=\frac{Fd+1}{d+1}, 
\end{equation}
where $f \in \left[ \frac{1}{d}, 1 \right]$ and $F \in \left[ \frac{1}{d^2}, 1 \right]$. Applications of the above relation include quantum communication~\cite{Yuan10}, quantum networks~\cite{Sasaki17}, quantum computing~\cite{Gottesman99}, quantum repeaters~\cite{Li19}, remote state preparation~\cite{Bennett01}, etc. 


Equation~\eqref{f_F_par} is valid for distinguishable particles only. 
For example, in Eq.~\eqref{f_F_par}, Putting $F=1$ yields $f=1$, meaning that unit teleportation fidelity is possible using distinguishable particles. But with indistinguishable particles, $f=1$ is not possible as shown by Das \textit{et al.}~\cite{Das20}.  Thus the relation between teleportation fidelity and singlet fraction in Eq.~\eqref{f_F_par} is no longer applicable for indistinguishable particles. 
Further, Eq.~\eqref{f_F_par} is applicable for particles with a single degree of freedom (DoF) only. 
However, the analysis for distinguishable particles with multiple DoFs is not as straightforward as those with single DoF. The situation becomes more complex for indistinguishable particles having multi-DoF entanglement, because the trace-out rule for indistinguishable particles with a single DoF~\cite{LFC16,LFC18} has to be reformulated for multi-DoFs~\cite{Das20,Paul21}.

 \subsection{\label{sec:level2}Results}
In this paper, we introduce the generalized teleportation fidelity and generalized singlet fraction and derive their relation applicable for both distinguishable and indistinguishable particles each having single or multiple DoFs. 
We also derive an upper bound for the generalized singlet fraction for distinguishable particles using the monogamy of singlet fraction~\cite{Kay09}.
This relation helps to characterize many composite systems based on their distinguishability, separability, presence of maximally entangled structure, and the number of DoFs. 

Further, our relationship unveils some counter-intuitive results. For example, $F=1$ means that the state is maximally entangled. However,  we propose an optical circuit to demonstrate that when the generalized singlet fraction is 1, it does not necessarily mean that the state is maximally entangled. We also analyze the optical circuit of~\cite{HHNL} to illustrate that the maximum value of the generalized singlet fraction can be greater than one. 

Finally, we show how using additional DoFs as ancilla qubits reduces the quantum resource requirements in quantum cryptographic protocols and characterize the states using the generalized singlet fraction. This results in new ways to characterize and analyze the states that have the same mathematical structure but different physical properties in terms of the number of particles and the number of DoFs.

\begin{figure*}[t!] 
\centering
\includegraphics[width=\textwidth]{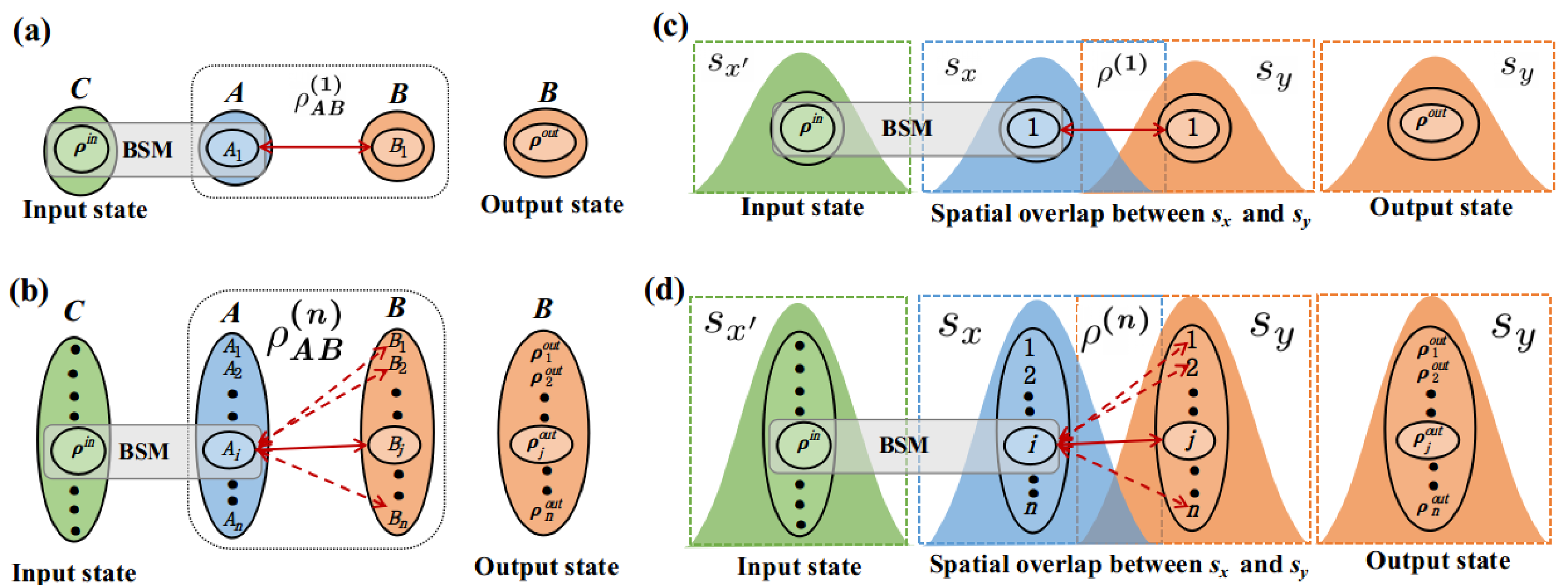}
\caption{Teleportation operation using Bell state measurement (BSM) for distinguishable and indistinguishable particles each having single or multiple DoFs. In (a) two distinguishable particles $A$ and $B$ each have a single DoF. In (b) two distinguishable particles $A$ and $B$, each having $n$ DoFs. In (c) two indistinguishable particles in spatial locations $s_x$ and $s_y$ each having a single DoF. In (d) two indistinguishable particles in spatial locations $s_x$ and $s_y$ each having $n$ DoFs.}
\label{teleport}
\end{figure*}

\section{Motivation for generalized definitions of teleportation fidelity and singlet fraction} \label{Gen_tel}

In this section,
first, we describe the standard teleportation protocol with distinguishable particles each having a single DoF. Then,  we increase the number of DoFs from single to multiple and explain how teleportation operations can be performed using distinguishable particles. Next, we move to the indistinguishable domain and discuss the teleportation operation using indistinguishable particles each having a single DoF. Finally, we present the most general version of the teleportation protocol with indistinguishable particles each having multiple DoFs that cover all the above cases. For each of the above scenarios, we discuss the applicability and drawbacks of the definition of teleportation fidelity (in Eq.~\eqref{Tf}), singlet fraction (in Eq.~\eqref{SFold}) and their relations (in Eq.~\eqref{f_F_par}).

To represent all the teleportation schemes, first, we define the general density matrix for two distinguishable and indistinguishable particles each having $n$ DoFs using the formalism of~\cite{LFC16,LFC18,Paul21}. 
\begin{definition} 
For two distinguishable particles $A$ and $B$ each having $n$ DoFs, the $i$-th and the $j$-th DoF are represented by $a_{i}$ (and $u_i$) and $b_{j}$ (and $v_j$) respectively where $i, j \in \mathcal{N} :=\lbrace 1, 2, \cdots , n \rbrace$.  Here, each DoF is $d$-dimensional whose eigenvalues are denoted by
$\mathbb{D}_{k}:=\lbrace D_{k_{1}}, D_{k_{2}}, \ldots, D_{k_{d}} \rbrace$ where $k \in \mathcal{N}$. Let $a_{(n)}$ denote the expression $a_{1} a_{2} \ldots a_{n} $ where  $a_{k} \in \mathbb{D}_k$ (and so $b_{(n)}, u_{(n)},$ and $v_{(n)}$ denote similar expressions).
 The density matrix of $A$ and $B$ is defined by~\cite{Paul21}
\begin{equation} \label{DMdis}
\begin{aligned}
\rho^{(n)}_{AB}:= \sum_{\substack{a_{(n)}, u_{(n)}  , \\ b_{(n)} ,v_{(n)}}} \kappa^{a_{(n)} u_{(n)} }_{ b_{(n)} v_{(n)}} \ket{a_{(n)}}_{A} \ket{b_{(n)}}_{B} \otimes \bra{ u_{(n)}}_{A}\bra{v_{(n)}}_{B},
\end{aligned}
\end{equation}
where the coefficients  $\kappa^{a_{(n)} u_{(n)} }_{ b_{(n)} v_{(n)}}$ satisfy the usual normalization condition.
\end{definition}

\begin{definition}
For two indistinguishable particles and for $p$ localized regions denoted as $ \mathcal{S} = \left\lbrace s_{1}, s_{2}, \cdots , s_{p} \right \rbrace $, the joint state is defined by~\cite{LFC16,LFC18,Paul21} 
\medmuskip=-1mu
\thinmuskip=-1mu
\thickmuskip=-1mu
\begin{equation} \label{DM}
\begin{aligned}
\rho^{(n)}:=  \sum_{\substack{\alpha,  \beta, \gamma, \delta  \in \mathcal{S}  \\ a_{(n)}, u_{(n)},b_{(n)},v_{(n)}}} \eta^{\mu} \kappa^{\alpha a_{(n)} \gamma u_{(n)} }_{\beta b_{(n)} \delta  v_{(n)}}   \ket{ \alpha a_{(n)}, \beta b_{(n)}}  \bra{\gamma u_{(n)}, \delta v_{(n)}}.
\end{aligned}
\end{equation}
\medmuskip=2mu
\thinmuskip=2mu
\thickmuskip=2mu
Here, $\mu$ denotes the sum of the parity of the cyclic permutations for all the DoFs where $\eta=+1$ for bosons and $\eta=-1$ for fermions. 
\end{definition}
If  
$\lbrace (\alpha=\beta ) \wedge ( a_{k}=b_{k} ) \rbrace$  or $ \lbrace (\gamma=\delta )   \wedge ( u_k=v_k ) ) \rbrace $
 for any  $k \in \mathcal{N}$, then $\eta=0$ for fermions due to the Pauli exclusion principle~\cite{Pauli25}. Other notations are the same as defined in Eq.~\eqref{DMdis} (for details see Appendix~\ref{Tr_out_rule}).

\subsection{Teleportation with two distinguishable particles each having a single DoF} \label{D1DoF}
Let's start with the standard model for teleportation with two distinguishable particles each having a single DoF as described in~\cite{QT93}. Suppose $\rho^{(1)}_{AB}$ denotes the joint state of two distinguishable particles $A$ and $B$, each having a single DoF denoted by $A_1$ and $B_1$ respectively, substituting $n=1$ in Eq.~\eqref{DMdis}.  Another particle $C$, also having a single DoF, has an unknown state in its only DoF denoted by $\rho^{in}$ which has to be teleported to $B$. For teleportation operation, Bell state measurement is performed jointly with the particles $C$ and $A$. After the Bell state measurement, the unknown state is teleported to $B$, denoted by $\rho^{out}$ as shown in Fig.~\ref{teleport} (a).

\subsection{Teleportation with two distinguishable particles each having multiple DoFs} \label{DnDOF}
 Now we move to a generalized teleportation with distinguishable particles where each particle has $n$ DoFs. Here, we consider two particles $A$ and $B$ each having $n$ DoFs, and their joint state is denoted by $\rho^{(n)}_{AB}$ as shown in Eq.~\eqref{DMdis}. Another particle $C$ also has $n$ DoFs and the unknown state $\rho^{in}$ that has to be teleported, exists in any one of the DoFs of $C$. Now in the previous case, only a single teleportation channel exists in between $A$ and $B$ as they have only a single DoF. But, as $A$ and $B$ have $n$ DoFs, then for each DoF in $A$, there exists $n$ teleportation channel in between $A$ to $B$. Thus for $n$ DoFs, total $n^{*}n=n^2$ teleportation channels are possible in between $A$ and $B$. 

Let us assume, we want to use the teleportation channel between the $i$th DoF of $A$ and the $j$th DoF of $B$. So, we have to apply Bell state measurement with the DoF of $C$ having the unknown state $\rho^{in}$ and the $i$-th DoF of the particle $A$. After Bell state measurement, the unknown state is teleported to all the DoFs of $B$ as shown in Fig.~\ref{teleport} (b). We denote the teleported state on the $j$-th DoF of the particle $B$ by $\rho^{out}_{j}$.  Note that, if there exists maximally entanglement between the $i$th DoF of $A$ and the $j$th DoF of $B$, then no entanglement will be present in all other teleportation channels involving the $i$th DoF of $A$ and the rest of the DoFs of $B$ due to monogamy of entanglement~\cite{CKW00}.  But, if there exists entanglement in all or some of the channels, then the unknown state will be teleported according to the amount of entanglement present in the channel. 

Now if we want to calculate the teleportation fidelity and singlet fraction for this case using Eq.~\eqref{Tf} and Eq.~\eqref{SFold} respectively, then it will not work. The reason is in Eq.~\eqref{Tf} and Eq.~\eqref{SFold}, the DoF information, the number of teleportation channels present, and using which teleportation is performed are missing.  Thus we need a new generalized formula to calculate teleportation fidelity and singlet fraction for this case, taking care of the above drawbacks. 
There can be another situation in which instead of a particular DoF, the whole particle with multiple DoFs has to be teleported from one location to another as shown in~\cite{Wang15}. In that case, the existing relations hold.

\subsection{Teleportation with two indistinguishable particles each having a single DoF} \label{ID1DOF}
 Next, we move from a distinguishable domain to an indistinguishable one. First, we consider teleportation using two indistinguishable particles each having a single DoF as shown in~\cite{LFC18}.  The teleportation operation using indistinguishable particles is almost the same as that of using distinguishable ones.  Consider two indistinguishable particles each having a single DoF with density matrix $\rho^{(1)}$ as defined in~Eq.~\eqref{DM} substituting $n=1$ in the left-hand side, located in the spatial regions $s_x$ and $s_y$ where $s_x, s_y \in \mathcal{S}$. The particle $C$, distinguishable from them, is located in the spatial region $s_{x^{\prime}} \in \mathcal{S}$ nearby $s_x$, having an unknown state $\rho^{in}$. 

 This unknown state has to be teleported in the spatial region $s_y$ from $s_x^{\prime}$. For that, Bell state measurement is performed jointly on the spatial regions $s_x$ and $s_x^{\prime}$. After this operation,  $\rho^{in}$ is teleported (probabilistically) to the spatial region $s_y$ as $\rho^{out}$ as shown in Fig.~\ref{teleport} (c). 
 
 In this case, the teleportation fidelity and the singlet fraction can be calculated as shown in Eq.~\eqref{Tf} and Eq.~\eqref{SFold} respectively. But their relation as shown in Eq.~\eqref{f_F_par} is no longer valid for indistinguishable particles. 
For example, in Eq.~\eqref{f_F_par}, setting $F = 1$ results in $f = 1$, indicating that unit teleportation fidelity is achievable using indistinguishable particles. But it was shown that with indistinguishable particles, $f = 1$ is not possible~\cite{Das20}. Thus we need a new relation between teleportation fidelity and singlet fraction for indistinguishable particles to take care of these confusions.

\subsection{Teleportation with two indistinguishable particles each having multiple DoFs} \label{IDnDOF}
Now we will consider the most generalized teleportation scheme using two indistinguishable particles each having $n$ DoFs. Note that, all the above cases can be derived from this case. Consider two indistinguishable particles with density matrix $\rho^{(n)}$ as defined in~Eq.~\eqref{DM}, located in the spatial regions $s_x$ and $s_y$ where $s_x, s_y \in \mathcal{S}$. The particle $C$, distinguishable from them, is located in the spatial region $s_{x^{\prime}} \in \mathcal{S}$ nearby to $s_x$, having an unknown state $\rho^{in}$ in one of its DoF. 
To teleport the unknown state from $s_x$ to  $s_y$,  Bell state measurement is performed on the $i$-th DoF of the spatial region $s_x$ and the DoF of $C$ having the unknown state $\rho^{in}$. After this operation,  $\rho^{in}$ is teleported (probabilistically) to all the DoFs of the spatial region $s_y$. We denote the teleported state of the $j$-th DoF of the spatial region $s_y$ by $\rho^{out}_{j}$ as shown in Fig.~\ref{teleport} (d). 

It is evident that the formulas for teleportation fidelity and singlet fraction and their relation are not valid in this case and we need to define generalized teleportation fidelity and generalized singlet fraction and derive their relation valid for both distinguishable or indistinguishable particles having single or multiple DoFs.

\section{Definitions of generalized teleportation fidelity and generalized singlet fraction and their relations } \label{Gen_TF_SF_NS}
In the previous section, we have described different types of teleportation schemes using distinguishable and indistinguishable particles. Here, we first define the generalized teleportation fidelity ($f_g$) and the generalized singlet fraction ($F^{(n)}_g$) for both distinguishable and indistinguishable particles each having multiple DoFs. Then we derive a generalized relation between $f_g$ and $F^{(n)}_g$ for a generalized noisy singlet state. Finally, we present the proof that our generalized relation applies to any quantum state.

\subsection{Generalized teleportation fidelity $f_{g}$} \label{gen_fg}

For any teleportation protocol, the motivation is to maximize the information transfer. So we define the generalized teleportation fidelity to capture the maximum fidelity over all possible channels. 
Here we define the generalized teleportation fidelity for indistinguishable particles, each having $n$ DoFs as discussed in Sec.~\ref{IDnDOF}.
\begin{definition}
The teleportation fidelity for $\rho^{in}_i$ using the $i$-th DoF of $A$ and the $j$-th DoF of $B$ as the channel is defined as
\begin{equation} \label{f_i_j}
f^{i}_{j}:=\text{Tr}\sqrt{\sqrt{\rho^{in}_i}\rho^{out}_{j}\sqrt{\rho^{in}_i}}.
\end{equation} 
As the goal of any teleportation protocol is to maximize the fidelity,
 the generalized  teleportation fidelity for the state $\rho^{(n)}$ is defined as
\begin{equation} \label{fg}
f_{g}:=\max_{ i,j  \in \mathcal{N}  } \left\lbrace  f^{i}_{j} \right\rbrace =\max_{ i,j  \in \mathcal{N}  } \left( \text{Tr}\sqrt{\sqrt{\rho^{in}_i}\rho^{out}_{j}\sqrt{\rho^{in}_i}} \right).
\end{equation}
\end{definition}
This definition of generalized teleportation fidelity applies to all the cases described in Fig.~\ref{teleport}.

\subsection{Generalized singlet fraction $ F^{(n)}_{g}$} \label{gen_Fn}
  For two distinguishable particles, say $A$ and $B$, each having $n$ DoFs as in Eq.~\eqref{DMdis}, if  $a_i$ is maximally entangled with $b_j$, then all the other pairs of the form $\lbrace a_i, b_k \rbrace$, $k\neq j$, as well as $\lbrace a_l, b_j \rbrace$, $l\neq i$, become separable due to monogamy of entanglement between DoFs~\cite{Camalet17,Camalet18}. For two indistinguishable particles each having $n$ DoFs as in Eq.~\eqref{DM}, it is possible to have all $\lbrace a_i, b_j \rbrace$ pairs maximally entangled as shown in~\cite{Paul21}. Thus the current definition of singlet fraction does not capture all such scenarios for distinguishable and indistinguishable particles. So the motivation to re-define generalized singlet fraction is to capture all the entangled or separable structures between two distinguishable or indistinguishable particles. 
 So we take the summation of all the singlet fractions possible for any particular DoF. Then we maximize that value for all DoFs of both the particles.
\begin{definition}
 For two distinguishable particles $A$ and $B$ with the joint state $\rho^{(n)}_{AB}$ as shown in~\eqref{DMdis}, we define the generalized singlet fraction as
\begin{equation} \label{SFGen}
 F^{(n)}_{g}:=\max \left\{  \max_{i} \lbrace F(i) \rbrace,   \max_{j} \lbrace F(j) \rbrace \right\},
\end{equation}
\medmuskip=2mu
\thinmuskip=2mu
\thickmuskip=2mu

where
\begin{equation}
\begin{aligned}
F(i):=& \max_{\psi_{a_{i}b_{{j}}}} { \sum^{n}_{j=1} P_{a_{i}b_{{j}}} }, \hspace{0.5cm}
F(j):=& \max_{\psi_{a_{i}b_{{j}}}} {  \sum^{n}_{i=1} P_{a_{i}b_{{j}}}  }, 
\end{aligned}
\end{equation}
and
\begin{equation}
\begin{aligned}
P_{a_{i}b_{j}}:=\braket{\psi_{a_{i}b_{j}} \mid \rho_{a_{i}b_{j}} |\psi_{a_{i}b_{j}}}, \hspace{0.5cm}
\rho_{a_{i}b_{j}}:= \text{Tr}_{a_{\bar{i}}{b_{\bar{j}}}} ( \rho^{(n)}_{AB} ).
\end{aligned}
\end{equation}
Here $a_{\bar{i}}=a_{1}a_{2} \cdots a_{i-1}a_{i+1} \cdots a_{n}$ and similar meaning for $b_{\bar{j}}$. 
In the terms $\max_{\psi_{a_{i}b_{{j}}}} \big\lbrace \sum^{n}_{j=1} P_{a_{i}b_{{j}}} \big\rbrace$, the $i$-th DoF of $A$ is kept fixed and $\ket{\psi_{a_{i}b_{{j}}}}$ spans all possible maximally entangled states between the $i$-th DoF of $A$ and the $j$-th DoF of $B$.
Similarly, for the other term, the $j$-th DoF of $B$ is kept fixed. 
\medmuskip=2mu
\thinmuskip=2mu
\thickmuskip=2mu 
\end{definition}
\begin{definition}
For two indistinguishable particles with a joint state $\rho^{(n)}$ as in Eq.~\eqref{DM}, each possessing $n$ DoFs, the generalized singlet fraction for two spatial regions $s_x, s_y \in \mathcal{S}$ (as depicted in Fig.~\ref{teleport} (b)) can be defined analogously using Eq.~\eqref{SFGen}, where $F(i)$ and $F(j)$ are defined by 
\begin{equation}
\begin{aligned}
F(i):=\max_{\psi_{s_{x_{i}}s_{y_{{j}}}}} { \sum^{n}_{j=1} P_{s_{x_{i}}s_{y_{{j}}}} }, \hspace{0.5cm}
 F(j):=\max_{\psi_{s_{x_{i}}s_{y_{{j}}}}} {  \sum^{n}_{i=1} P_{s_{x_{i}}s_{y_{{j}}}}  }, 
 \end{aligned}
\end{equation}
and $P_{s_{x_{i}}s_{y_{{j}}}}$ and $ \rho_{s_{x_{i}}s_{y_{{j}}}}$ are defined by
\frenchspacing
 \medmuskip=-2mu
\thinmuskip=-2mu
\thickmuskip=-2mu 
 \begin{equation}
\begin{aligned}
 P_{s_{x_{i}}s_{y_{{j}}}}:=\braket{\psi_{s_{x_{i}}s_{y_{{j}}}}|\rho_{s_{x_{i}}s_{y_{{j}}}} |\psi_{s_{x_{i}}s_{y_{{j}}}}}, \hspace{0.3cm}
 \rho_{s_{x_{i}}s_{y_{{j}}}}:= \text{Tr}_{s_{x_{\bar{i}}}s_{y_{\bar{j}}}} ( \rho^{(n)} ).
\end{aligned}
\end{equation}
\frenchspacing
 \medmuskip=2mu
\thinmuskip=2mu
\thickmuskip=2mu 
Here $\ket{\psi_{s_{x_{i}}s_{y_{{j}}}}}$ spans all possible maximally entangled states between the $i$-th DoF of $s_x$ and the $j$-th DoF of $s_y$ and $s_x, s_y \in  \mathcal{S} $, following the DoF trace-out rule in Appendix~\ref{Parti_DoF_Tr}.
\end{definition} 
\subsection{Relation between  generalized teleportation fidelity $f_{g}$ and generalized singlet fraction  $ F^{(n)}_{g} $ for generalized Werner states} \label{Rel_fg_Fn_NS}
\begin{definition} \label{D_twopara}
Consider a two-parameter family of states for two particles each having $n$ DoFs defined as
\begin{equation} \label{twopara}
\rho^{(n)}_{p}:= p \mathbb{P}^{(n)} + (1-p) \frac{\mathbb{I}^{n} \otimes \mathbb{I}^{n} }{d^{2n}}, \hspace{1cm} 0 \leq p \leq 1,
\end{equation} 
\noindent
where for two distinguishable particles $A$ and $B$, the state $\rho^{(n)}_{p}$ is an arbitrary $\rho^{(n)}_{AB}$ and  $ \mathbb{P}^{(n)}$ is a special form of $ \rho^{(n)}_{AB}$ such that for every $a_i$, there exists at least one $b_j$ with $\mathbb{P}_{a_{i}b_{j}}=\text{Tr}_{a_{\bar{i}}b_{\bar{j}}} \left( \mathbb{P}^{(n)} \right)$ as $d$-dimensional maximally entangled. Similarly, for two indistinguishable particles, $\rho^{(n)}_{p}$ is an arbitrary  $\rho^{(n)}$ and $\mathbb{P}^{(n)} $ is a special form of $\rho^{(n)}$  such that $\mathbb{P}_{s_{x_{i}}s_{y_{{j}}}} = \text{Tr}_{s_{x_{\bar{i}}}s_{y_{{\bar{j}}}}} \left( \mathbb{P}^{(n)} \right)$ is maximally entangled. Thus this equation is applicable for both distinguishable and indistinguishable particles.
 \end{definition}
This type of state is possible using indistinguishable particles as shown in~\cite{Paul21}. But for distinguishable particles, such type of state is not possible as they obey monogamy of entanglement~\cite{Camalet17}.
 Now we derive the relation between generalized teleportation fidelity $f_{g}$ and generalized singlet fraction  $ F^{(n)}_{g} $ for generalized Werner states as defined in Eq.~\eqref{twopara}.

First, we calculate the generalized teleportation fidelity for the state $\rho^{(n)}_{p}$ and
or that we calculate the same separately for both $\mathbb{P}^{(n)}$ and  completely random noise $(\mathbb{I}^{n} \otimes \mathbb{I}^{n} )/d^{2n}$. 
We denote $f_{max}$ to be the value of $f_{g}$ for the state $\mathbb{P}^{(n)} $.
\begin{Lemma} \label{L_genTF}
    The generalized  teleportation fidelity applicable  for both distinguishable and indistinguishable particles for the state in Eq.~\eqref{twopara} is
\begin{equation} \label{fgparam}
f_{g} = p f_{max} + (1-p)\frac{1}{d},
\end{equation}
where $f_{max}=1$ for distinguishable particles and $f_{max} < 1$ for indistinguishable particles.
\end{Lemma}
\begin{proof}

The difference between $f_{g}$ and $f_{max}$ is that the first one is the maximum value of $f^{i}_{j}$ for an arbitrary state $\rho^{(n)}_{AB}$  or $\rho^{(n)}$  but the second one is the maximum value of $f^{i}_{j}$ for the special state $\mathbb{P}^{(n)} $.
 For distinguishable particles, $\max_{(i,j)} \lbrace f^{i}_{j} \rbrace=1$  occurs for the pair $(i, j)$ such that $\mathbb{P}_{a_{i}b_{{j}}}$ is maximally entangled. For indistinguishable particles, $f_{max} < 1$ due to the \textit{no-go} theorem of~\cite{Das20}.  The value of $f_g$ for  $(\mathbb{I}^{n} \otimes \mathbb{I}^{n} )/d^{2n}$  is $1/d$  for both distinguishable or indistinguishable particles as  $\rho^{out}=\mathbb{I}/d$ independent of the initial state $\rho^{in}$ before teleportation. 
Thus we get the generalized teleportation fidelity applicable for both distinguishable and indistinguishable particles for the state in Eq.~\eqref{twopara} as shown in Eq.~\eqref{fgparam}.
\end{proof}
Next, we calculate the generalized singlet fraction for the state $\rho^{(n)}_{p}$. Let  $F^{(n)}_{max}$ be the value of $F^{(n)}_{g}$ for the state $\mathbb{P}^{(n)}$. 
\begin{Lemma} \label{L_GenSF}
    The generalized singlet fraction  for both distinguishable and indistinguishable particles for the state in Eq.~\eqref{twopara} is
    \begin{equation} \label{Fn}
  F^{(n)}_{g} = p F^{(n)}_{max} + (1-p)\frac{n}{d^{2}}.
 \end{equation}
\end{Lemma}
\begin{proof}
We know that distinguishable particles obey the monogamy of entanglement. So, let us fix a particular $i$, then  
$P_{a_{i}b_{j}}=\braket{\psi_{a_{i}b_{j}} \mid \mathbb{P}^{(n)}_{a_{i}b_{j}} \mid \psi_{a_{i}b_{j}}}$ would be 1 for a particular $j$, say $j^{\prime}$, as $\mathbb{P}^{(n)}_{a_{i}b_{j^{\prime}}}$ is maximally entangled. When $j \neq j^{\prime}$, we get $P_{a_{i}b_{j}}=1/d$ as $\mathbb{P}^{(n)}_{a_{i}b_{j}}$ is separable, where $i, j , j^{\prime} \in \mathcal{N}$. So, for the rest of the $(n-1)$ DoFs, we get $P_{a_{i}b_{j}}=1/d$. Thus the value of $F^{(n)}_{max}$ for distinguishable particles is $ ( 1+\frac{(n-1)}{d} ) $.
It is proved in~\cite{Paul21} that two indistinguishable particles, each having multiple DoFs do not obey the monogamy of entanglement. So, if we fix any particular $i$, then $P_{s_{x_{i}}s_{y_{{j}}}}$ can be 1 for all $j$, as $\mathbb{P}_{s_{x_{i}}s_{y_{{j}}}}$ can be maximally entangled for all $j$. So, for indistinguishable particles $ \max_{i} \lbrace F(i) \rbrace$  is $n$, and similarly  $  \max_{j} \lbrace F(j) \rbrace=n$, leading to $F^{(n)}_{max}=n$.
For  $\mathbb{I}^{n} \otimes \mathbb{I}^{n}/d^{2n}$, we  get $P_{a_{i}b_{j}}=P_{s_{x_{i}}s_{y_{{j}}}}=1/d^{2}$ for all $i$ and $j$. So,  $F^{(n)}_{g}$ is $n/d^2$ which is the same for distinguishable or indistinguishable particles.
Thus, we get the value of $F^{(n)}_{g}$ for Eq.~\eqref{twopara} as given in Eq.~\eqref{Fn}.
\end{proof} 

\begin{Theorem} \label{Th:sp}
   The relation between  generalized teleportation fidelity $f_{g}$ as shown Lemma~\ref{L_genTF} in and generalized singlet fraction  $ F^{(n)}_{g} $ as shown Lemma~\ref{L_GenSF} for generalized Werner states $\rho^{(n)}_{p}$ as defined in Definition~\ref{D_twopara} is given by
    \begin{equation} \label{fandF}
f_{g}=\frac{\left( F^{(n)}_{g} - \frac{n}{d^{2}} \right) \left( f_{max} -\frac{1}{d}\right)  }{\left( F^{(n)}_{max}-\frac{n}{d^{2}} \right)}  + \frac{1}{d},
\end{equation}
where $f_{g} \in \left[\frac{1}{d}, f_{max} \right] $ and $F^{(n)}_{g} \in \left[ \frac{n}{d^{2}}, F^{(n)}_{max} \right] $. 
\end{Theorem}
\begin{proof}
Using Eq.~\eqref{fgparam} and Eq.~\eqref{Fn}, we get the relation between the generalized teleportation fidelity and the generalized singlet fraction as shown in Eq.~\eqref{fandF}.
Whether the particles are distinguishable or indistinguishable, the same Eq.~\eqref{fandF} holds, only the values of $f_{max}$ and $F^{(n)}_{max}$ vary.
For particles having single DoF, i.e., $n=1$, Eq.~\eqref{fandF} reduces to Eq.~\eqref{f_F_par}.
\end{proof}
\subsection{Proof of generalized relation between the generalized teleportation fidelity and the generalized singlet fraction for any state} \label{Rel_fg_Fn}

 This section will provide detailed proof of the generalized relation between generalized teleportation fidelity and the generalized singlet fraction for any state. Note that the approach is similar to the original proof of Eq.~\eqref{f_F_par} (details are given in Appendix~\ref{HorodeckiProof}), but the detailed calculation is non-trivial as it involves the presence of multiple DoFs and indistinguishability. 
To prove this, we must first establish the following definitions and lemmas.

Nosrati \textit{et al.}~\cite{Nosrati20} have shown a maximally entangled state using two indistinguishable particles in their spin DoF. Here, we will define a maximally entangled state using two indistinguishable particles each having a single DoF.
\begin{definition} \label{D_MES}
Let us consider two indistinguishable particles each having a single DoF, having two localized regions denoted by $ \mathcal{S} = \left\lbrace s_{1}, s_{2} \right \rbrace $. Suppose the DoF is $d$-dimensional whose eigenvalues are denoted by
$\mathbb{D}:=\lbrace D_{1}, D_{2}, \ldots, D_{d} \rbrace$. The general maximally entangled state of two such particles can be defined by
\begin{equation} \label{State2pId}
\begin{aligned}
\ket{\Psi^{(2)}}_{\text{MES}}:= \dfrac{1}{\sqrt{d}}\sum_{\alpha, \beta \in \mathcal{S}, a \in \mathbb{D} }\eta^{\nu} \kappa^{\alpha \beta
}_{aa} \ket{\alpha a, \beta  a},
\end{aligned}
\end{equation}
where, $  \alpha , \beta$ ranges over $ \mathcal{S} $; $a$ ranges over $ \mathbb{D}$. Here, $\nu$ denotes the parity of the cyclic permutations for the DoF where $\eta=+1$ for bosons and $\eta=-1$ for fermions. Now we define the density matrix corresponding to the maximally entangled state as 
 $ \mathbb{P}^{(2)}_{\text{MES}} = \ket{\Psi^{(2)}}_{\text{MES}} \bra{\Psi^{(2)}}_{\text{MES}}$. 
\end{definition}

Horodecki \textit{et al.}~\cite{Horodecki99} have proved the isomorphism between the channels and the states for distinguishable particles each having a single DoF only. Here, we will generalize this proof for both distinguishable and indistinguishable particles each having single or multiple DoFs. Note that, this proof may be trivial for distinguishable particles each having multiple DoFs, but it is not so for indistinguishable particles each having single or multiple DoFs.
      
 \begin{Lemma} \label{rho_Ch_Indis}
     There exists an isomorphism between channels and states for both distinguishable and indistinguishable particles each having single or multiple DoFs.
 \end{Lemma}
\begin{proof}
First, we will prove the isomorphism between channels and states for indistinguishable particles each having a single DoF.   For a completely positive trace preserving (CPTP) map $\Lambda$, a state made of two indistinguishable particles denoted by $\rho^{(2)}_{\Lambda}$, can be attributed by sending half of the particles without knowing which particles are being sent, through the channel $\Lambda$ as 
 \begin{equation} \label{rho_lambda}
 \rho^{(2)}_{\Lambda} = \left( \mathbb{I} \otimes \Lambda \right)  \mathbb{P}^{(2)}_{\text{MES}} + \eta \left(  \Lambda \otimes \mathbb{I}  \right)  \mathbb{P}^{(2)}_{\text{MES}},
 \end{equation}
 where $\mathbb{P}^{(2)}_{\text{MES}}$ is defined in Definition~\ref{D_MES} and $\eta=+1$ for bosons and $\eta=-1$ for fermions.
 For the state in Eq.~\eqref{rho_lambda}, if we trace-out one of the particles, we get the same state as when we trace-out one of the particles from $\mathbb{P}^{(2)}_{\text{MES}}$, i.e., a completely mixed state.
 
Now for the second part, i.e., state-to-channel isomorphism, we consider a given density matrix  $\rho^{(2)}$ made of indistinguishable particles, we have the spectral decomposition as
\begin{equation} \label{Spectral_decom}
\rho^{(2)} = \sum^{d^2}_{k=1} p_{k} \ket{\psi_{k}}\bra{\psi_{k}},
\end{equation}
where each term is represented, for $k=1$, we have $\ket{\psi^{(2)}_{1}} $ as
\begin{equation}
\ket{\psi^{(2)}_{1}} = \sum_{\alpha, \beta \in \mathcal{S}, a, b \in \mathbb{D} }\eta^{\nu} \kappa^{\alpha a}_{\beta  b} \ket{\alpha a, \beta  b}. 
\end{equation}
Further, we can represent $\ket{\psi^{(2)}_{1}} $ as
\begin{equation}
\ket{\psi^{(2)}_{1}} = \left( \mathbb{I} \otimes V^{(1)}_{\varsigma_1,\sigma_1} \right) \ket{\Psi^{(2)}}_{\text{MES}}
\end{equation}
where $\braket{  \alpha a\mid V^{(1)}_{\varsigma_1,\sigma_1} \mid \beta b}=\sqrt{d} \kappa^{\alpha a}_{\beta  b} $. Similarly, we can define $V^{(k)}_{\varsigma_1,\sigma_1}$  for $k \in \lbrace 1, 2, \ldots, d^2 \rbrace$
\begin{equation}
\begin{aligned}
\rho^{(2)} =& \sum^{d^2}_{k=1} p_{k} \left( \mathbb{I} \otimes V^{(1)}_{\varsigma_1,\sigma_1} \right) \mathbb{P}^{(2)}_{\text{MES}} \left( \mathbb{I} \otimes V^{(1)^{\dagger}}_{\varsigma_1,\sigma_1} \right) \\ =&  \left( \mathbb{I} \otimes \Lambda_{\rho^{(2)}} \right)  \mathbb{P}^{(2)}_{\text{MES}},
\end{aligned}
\end{equation}
where
\begin{equation*}
\Lambda_{\rho^{(2)}} (\varrho^{(2)}) =  \sum_{k} p_{k} V^{(1)}_{\varsigma_1,\sigma_1}  \varrho^{(2)} V^{(1)^{\dagger}}_{\varsigma_1,\sigma_1}.
\end{equation*}
Further, it can be shown easily that for two channels $\Lambda_{\rho^{(2)}}$ and $\Lambda^{\prime}_{\rho^{(2)}}$ that produces the same state $\rho^{(2)}$, we have  
\begin{equation}
\left[ \mathbb{I} \otimes \left(\Lambda_{\rho^{(2)}} -\Lambda^{\prime}_{\rho^{(2)}} \right)\right] = 0.
\end{equation}

This proves that the channel $\Lambda_{\rho^{(2)}}$ is unique for the density matrix $\rho^{(2)}$ and hence the isomorphism between the channels and the states made of indistinguishable particles having a single DoF. Since the entire proof is not dependent on the number of DoFs of each particle, we can get the same isomorphism between the channels and the states made of distinguishable or indistinguishable particles each having multiple DoFs. 
\end{proof}



Before going to the next lemma, we will define \textit{deformations}~\cite{Piccolini21,Piccolini23} that is analogous to unitary transformations for distinguishable particles. 
 \begin{definition} \label{D_deform}
 For two indistinguishable particles as defined in Eq.~\eqref{State2pId}, the deformation $\mathcal{D}^{(2)}_{\vec{\varsigma}, \vec{\sigma}}$  is defined by 
 \begin{equation}
 \begin{aligned}
& \mathcal{D}^{(2)}_{\vec{\varsigma}, \vec{\sigma}} \ket{\Psi^{(2)}}_{\text{MES}} = \left( U^{(1)}_{\varsigma_1,\sigma_1} \otimes U^{(1)}_{\varsigma_2,\sigma_2}  \right)  \ket{\Psi^{(2)}}_{\text{MES}} \\ &=\dfrac{1}{\sqrt{d}}
 \sum_{\substack{ \alpha, \beta \in \mathcal{S}, \\ a, b \in \mathbb{D}, a\neq b, \vec{\varsigma} } }\mid \braket{ \sigma_1 \mid \alpha } \braket{ \sigma_2 \mid \beta } \mid\eta^{\nu} \ket{ U^{(1)}_{\varsigma_1,\sigma_1} \alpha a, U^{(1)}_{\varsigma_2,\sigma_2} \beta  a} 
 \end{aligned}
 \end{equation}
Here, the elements $\varsigma_i$ in $\vec{\varsigma}\in \lbrace \varsigma_{1}, \varsigma_{2} \rbrace$ identify the
type of transformation represented by the single particle unitary operator $U^{(1)}_{\varsigma_i,\sigma_i}$ where $i \in \lbrace 1, 2 \rbrace $ and $\sigma_i$ in $\vec{\sigma} \in \lbrace \sigma_1, \sigma_2 \rbrace$ denotes its region of action. Other notations have the same meaning as defined in Eq.~\eqref{State2pId}. 
\end{definition}
The deformations for $N$ indistinguishable particles can be found in~\cite{Piccolini21,Piccolini23}.
 
 \begin{Lemma} \label{rho_rho_NS_Indis}
     Any state $\rho^{(n)}$ under $\mathcal{D} \otimes \mathcal{D}^{*}$ transformation becomes a  noisy singlet $\rho^{(n)}_{p}$.
 \end{Lemma}
\begin{proof}
 The proof of this statement for indistinguishable particles is similar to that of distinguishable ones~\cite{Horodecki99}. Suppose we have a state $\rho^{(n)}$ made of indistinguishable particles. If we apply a twirling operation, i.e., $\mathcal{D} \otimes \mathcal{D}^{*}$ transformation, analogous to $U \otimes U^{*}$ ($U$ means unitary)  transformation for distinguishable particles as shown in Definition~\ref{D_deform}, then all the singlet states will remain singlet and all the other non-singlet ones will be averaged out. So the resultant state will be $\rho^{(n)}_{p}$. The reason behind this is that under $\mathcal{D} \otimes \mathcal{D}^{*}$ transformation, one cannot increase the entanglement, only reduction of entanglement is possible. So for non-singlet states, application of these deformations many times and taking the average of the results in a non-entangled state. Thus to prove the above statement, we have to show that  $\mathbb{P}^{(n)}_{\text{MES}}$ is invariant under $\mathcal{D} \otimes \mathcal{D}^{*}$ transformation. 


Let us define the deformation $\mathcal{D}$ (omitting the superscripts and subscripts for simplicity) for indistinguishable particles analogous to $U$ for distinguishable particles. Now we can have the following
\begin{equation}
\begin{aligned}
 \mathcal{D} \otimes \mathcal{D}^{*} \mathbb{P}^{(n)}_{\text{MES}} \mathcal{D}^{\dagger} \otimes \mathcal{D}^{* \dagger} =&  \mathbb{I} \otimes \mathcal{D}^{*} \mathcal{D}^{T}  \mathbb{P}^{(n)}_{\text{MES}} \mathbb{I} \otimes \left( \mathcal{D}^{*} \mathcal{D}^{T}\right)^{\dagger} \\=& \mathbb{P}^{(n)}_{\text{MES}}.
\end{aligned}
\end{equation}
Here we have used the identity $\left( \mathcal{D} \otimes \mathbb{I} \right) \ket{\Psi^{(n)}}_{\text{MES}} = \left(\mathbb{I}  \otimes \mathcal{D}^{T} \right)\ket{\Psi^{(n)}}_{\text{MES}}$ that can be proved for indistinguishable particles using similar methods as shown in~\cite{Jozsa94} where $T$ denotes transpose operation. 

The identity operator is indeed invariant under $\mathcal{D} \otimes \mathcal{D}^{*}$ transformation. Using similar arguments we can prove that the noisy singlet as proposed in~\eqref{f_F_par} for distinguishable and indistinguishable particles having multiple DoFs are invariant under $\mathcal{D} \otimes \mathcal{D}^{*}$ transformation.

\end{proof}

Let us first define the depolarizing channel below to prove the next lemma.
\begin{definition} \label{D_depo}
We define the depolarizing channel for the parameter $p$ as
\begin{equation} \label{depola}
\Lambda^{\text{dep}}_{p} (\sigma) := p \sigma + (1-p)\frac{\mathbb{I}}{d}, \hspace{1cm} 0 \leq p \leq 1,
\end{equation}
where $\sigma$ is the state acting on $d$-dimensional Hilbert space. 
\end{definition}

\begin{Lemma} \label{Dep_rhoNS_Indis}
    There exists an equivalence between the depolarizing channel $\Lambda^{\text{dep}}_{p}$  and noisy singlet $\rho^{(n)}_{p}$ for both distinguishable and indistinguishable particles
\end{Lemma}
\begin{proof}
Depolarizing channel as defined in Definition~\ref{D_depo} signifies that with probability $p$, it does not affect the input state and with probability $(1-p)$, it completely randomizes the input.
Now if we apply the channel in Eq.~\eqref{depola} to the half of the particles of the singlet state $\mathbb{P}^{(n)}$ as shown in the first term in right-hand side of Eq.~\eqref{twopara}, then we get the noisy singlet state $\rho^{(n)}_{p}$ as shown in the left-hand side of Eq.~\eqref{twopara}. Thus we have a full equivalence between the depolarizing channel and the noisy singlet state as
\begin{equation}
\Lambda^{\text{dep}}_{p} \equiv \rho^{(n)}_{p}.
\end{equation}
The above equivalence holds for both distinguishable and indistinguishable particles. 
\end{proof}
\begin{Lemma}
    Any channel $\Lambda$ subjected to a twirling process denoted by $\mathcal{T}$ becomes a depolarizing channel denoted by $\Lambda^{\text{dep}}_{p}$.
\end{Lemma}
\begin{proof}

This equivalence can be proved by the following arguments. In Sec.~\ref{rho_Ch_Indis}, we have proved the isomorphism between any state made of indistinguishable particles $\rho^{(n)}$ and a channel $\Lambda$. Thus we can write
 \begin{equation} 
 \begin{aligned}
   \rho^{(n)} & \equiv \Lambda,
   \end{aligned}
 \end{equation}
 If we apply the twirling operation on both sides we get
 \begin{equation} \label{Eq_rhoNS_TwCh_Indis}
 \begin{aligned} 
 \mathcal{T} \left( \rho^{(n)} \right)  & \equiv  \mathcal{T} \left(  \Lambda \right), \\
 \text{or}\\
 \rho^{(n)}_{p} & \equiv  \mathcal{T} \left(  \Lambda \right),
  \end{aligned}
 \end{equation}
 as any state $\rho^{(n)}$ under twirling operations becomes a  noisy singlet as shown in Sec.~\ref{rho_rho_NS_Indis}. Again from Sec.~\ref{Dep_rhoNS_Indis}, we can write that
  \begin{equation} \label{Eq_rhoNS_Ch_Indis}
 \begin{aligned}
  \rho^{(n)}_{p} \equiv \Lambda^{\text{dep}}_{p}.
 \end{aligned}
 \end{equation}
From Eq.~\eqref{Eq_rhoNS_TwCh_Indis} and~\eqref{Eq_rhoNS_Ch_Indis}, we can write
 \begin{equation}
  \mathcal{T} \left(  \Lambda \right) = \Lambda^{\text{dep}}_{p}.
 \end{equation}
Thus any channel under the twirling process becomes a depolarizing channel.
\end{proof}
\begin{Lemma}
    Relation between  $f_{g}$ and  $ F^{(n)}_{g} $ for any depolarizing channel $\Lambda^{\text{dep}}_{p}$ is given by
    \begin{equation} \label{fandF_Lam_dep}
f_{g} \left(\Lambda^{\text{dep}}_{p} \right) =\frac{\left( F^{(n)}_{g} \left(\Lambda^{\text{dep}}_{p} \right) - \frac{n}{d^{2}} \right) \left( f_{max}  -\frac{1}{d}\right)  }{\left( F^{(n)}_{max}-\frac{n}{d^{2}} \right)}  + \frac{1}{d}.
\end{equation}
\end{Lemma}
\begin{proof}
    We can denote the generalized singlet teleportation fidelity and generalized singlet fraction for channel $\Lambda$ by $f_{g} (\Lambda)$ and $F_{g}(\Lambda)$ respectively. In Eq.~\eqref{fandF}, we have defined the relation for the noisy singlet state $\rho^{(n)}_{p}$. Now as in Eq.~\eqref{Eq_rhoNS_Ch_Indis}, it is shown that the channel $\Lambda^{\text{dep}}_{p}$ is equivalent to the noisy singlet state $\rho^{(n)}_{p}$, we can write Eq.~\eqref{fandF} as Eq.~\eqref{fandF_Lam_dep}. This completes the proof.
\end{proof}
Now let us define generalized teleportation fidelity for any channel before going to the next lemma.
\begin{definition} \label{Def_fg_channel}
 The generalized teleportation fidelity as defined in Eq.~\eqref{fg} for any channel $\Lambda$ is similar to as defined in~\cite[Eq. (10)]{Horodecki99} is defined by
\begin{equation} \label{fg_channel}
f_{g} \left( \Lambda \right) := \int d\phi \braket{\phi\mid \Lambda \left( \ket{\phi}\bra{\phi} \right) \mid \phi},
\end{equation}
where the integration is performed with respect to the uniform distribution $d \phi$ over all input pure states.
\end{definition}

\begin{Lemma}
 There exists  invariance of $f_{g}$ and $F^{(n)}_{g}$ for twirling channels
\end{Lemma}
\begin{proof}
Here, we first introduce the operations equivalent to $\mathcal{D} \otimes \mathcal{D}^{*}$ twirling of states using distinguishable or indistinguishable particles with multiple DoFs. For any given channel $\Lambda$, we can construct the twirling of channels in the following way. Alice performs a random deformation $\mathcal{D}$ to the incoming particles. Then she sends those particles through the channel $\Lambda$ and informs Bob about the deformation. Now, Bob after receiving the particles applies an inverse transformation $\mathcal{D}^{* \dagger}$. Note that, twirling channels are independent of both particle indistinguishability and the number of DoFs.

 Now, we  prove first that under the twirling operation $\mathcal{T}$ on a channel $\Lambda$, the value of $f_{g}$ remains invariant, i.e.,
 \begin{equation}
 f_{g} \left( \mathcal{T} \left( \Lambda\right)  \right) = f_{g} \left(  \Lambda \right).
 \end{equation}

It is interesting to note that the definition of generalized teleportation fidelity for the channel (as defined in definition~\ref{Def_fg_channel}) doesn't depend either on the distinguishability of states that are being used or the number of DoFs present in the states of each particle. Thus the formula for $f_{g}$ for the channel $\Lambda$ will be the same as if we apply a twirling operation in the channel such as
\begin{equation}
f_{g} \left( \Lambda \right) = f_{g} \left( \mathcal{T} \left( \Lambda \right) \right).
\end{equation}
The formal proof for this would be similar to the proof given in~\cite[Eq. (20)]{Horodecki99}, where the unitary operations would be replaced with deformation (as defined in Definition~\ref{D_deform}) operations for indistinguishable particles. Thus the generalized channel fidelity $f_{g} $ is invariant under twirling operation. 

Next we will prove that under the twirling operation $\mathcal{T}$ on a channel $\Lambda$, the value of $F^{(n)}_{g}$ remains invariant, i.e.,
 \begin{equation}
 F^{(n)}_{g} \left( \mathcal{T} \left( \Lambda\right)  \right) = F^{(n)}_{g} \left(  \Lambda \right).
 \end{equation}
 
As we know that the channel $\Lambda$ under the twirling operation becomes a depolarizing channel, $\Lambda^{\text{dep}}_{p}$,
 we have 
\begin{equation}
F^{(n)}_{g} \left( \mathcal{T} \left( \Lambda\right)  \right) = F^{(n)}_{g} \left( \Lambda^{\text{dep}}_{p} \right).
\end{equation}
Again the depolarizing channel is equivalent to the noisy singlet states 
 and thus we have
\begin{equation}
F^{(n)}_{g} \left( \Lambda^{\text{dep}}_{p}\right)  \equiv F^{(n)}_{g} \left( \rho^{(n)}_{p} \right).
\end{equation}
Since noisy singlet states can be created by twirling operation on any state, we have
\begin{equation}
F^{(n)}_{g} \left( \rho^{(n)}_{p} \right) = F^{(n)}_{g} \left( \mathcal{T} \left( \rho^{(n)} \right)  \right). 
\end{equation}
 As $F^{(n)}_{g}$ is invariant under the twirling operation under states, we can write
 \begin{equation}
 F^{(n)}_{g} \left( \mathcal{T} \left( \rho^{(n)} \right)  \right) =F^{(n)}_{g} \left(  \rho^{(n)} \right).
 \end{equation}
Finally,  from the isomorphism between the channel and the states, we can write
\begin{equation}
F^{(n)}_{g} \left(  \rho^{(n)} \right) = F^{(n)}_{g} \left(  \Lambda \right). 
\end{equation}

Thus we can say that the generalized relation between the generalized channel fidelity and the generalized singlet fraction in Eq.~\eqref{fandF} is true for any channel $\Lambda$, i.e., 

\begin{equation} \label{fandF_ch}
f_{g} \left( \Lambda \right) =\frac{\left( F^{(n)}_{g} \left( \Lambda\right) - \frac{n}{d^{2}} \right) \left( f_{max} -\frac{1}{d}\right)  }{\left( F^{(n)}_{max}-\frac{n}{d^{2}} \right)}  + \frac{1}{d}.
\end{equation}
Logically we can say that Eq.~\eqref{fandF_ch} is true for depolarizing channel and further $f_{g}$ and $F^{(n)}_{g}$ are invariant under twirling, thus it is true for any channel.  
\end{proof}
Now we will prove the following theorem using the above lemmas.
\begin{Theorem} \label{Proof4fandF}
    The generalized relation between the generalized teleportation fidelity $f_{g}$ and the generalized singlet fraction  $ F^{(n)}_{g} $ as proved in Theorem~\ref{Th:sp} for the generalized Werner states, is also applicable for any state $\rho^{(n)}$.
\end{Theorem}
\begin{proof}

Since there is an isomorphism between channels and states, as Eq.~\eqref{fandF_ch} is true for any channel,  it must be true for any state. Thus we have 
\begin{equation} \label{fandF_ch_indis}
f_{g} \left( \rho^{(n)} \right) =\frac{\left( F^{(n)}_{g} \left( \rho^{(n)} \right) - \frac{n}{d^{2}} \right) \left( f_{max} -\frac{1}{d}\right)  }{\left( F^{(n)}_{max}-\frac{n}{d^{2}} \right)}  + \frac{1}{d}.
\end{equation}
This completes the proof that the generalized relation is valid between  $f_{g}$ and  $ F^{(n)}_{g} $ for all states made of indistinguishable particles.
\end{proof}

\section{upper bound on the generalized singlet fraction for distinguishable particles} \label{UppBound}
As monogamy of entanglement does not hold for indistinguishable particles~\cite{Paul21}, it is trivial to calculate the upper bound on generalized singlet fraction for distinguishable particles which equals the number of DoFs present in each particle. However, for distinguishable particles, the same is not trivial.
In this section, we derive the upper bound for generalized singlet fraction for distinguishable particles using the monogamy of singlet fraction~\cite{Kay09} as shown below.
\begin{figure*}[t!] 
\centering
\includegraphics[width=0.8\textwidth]{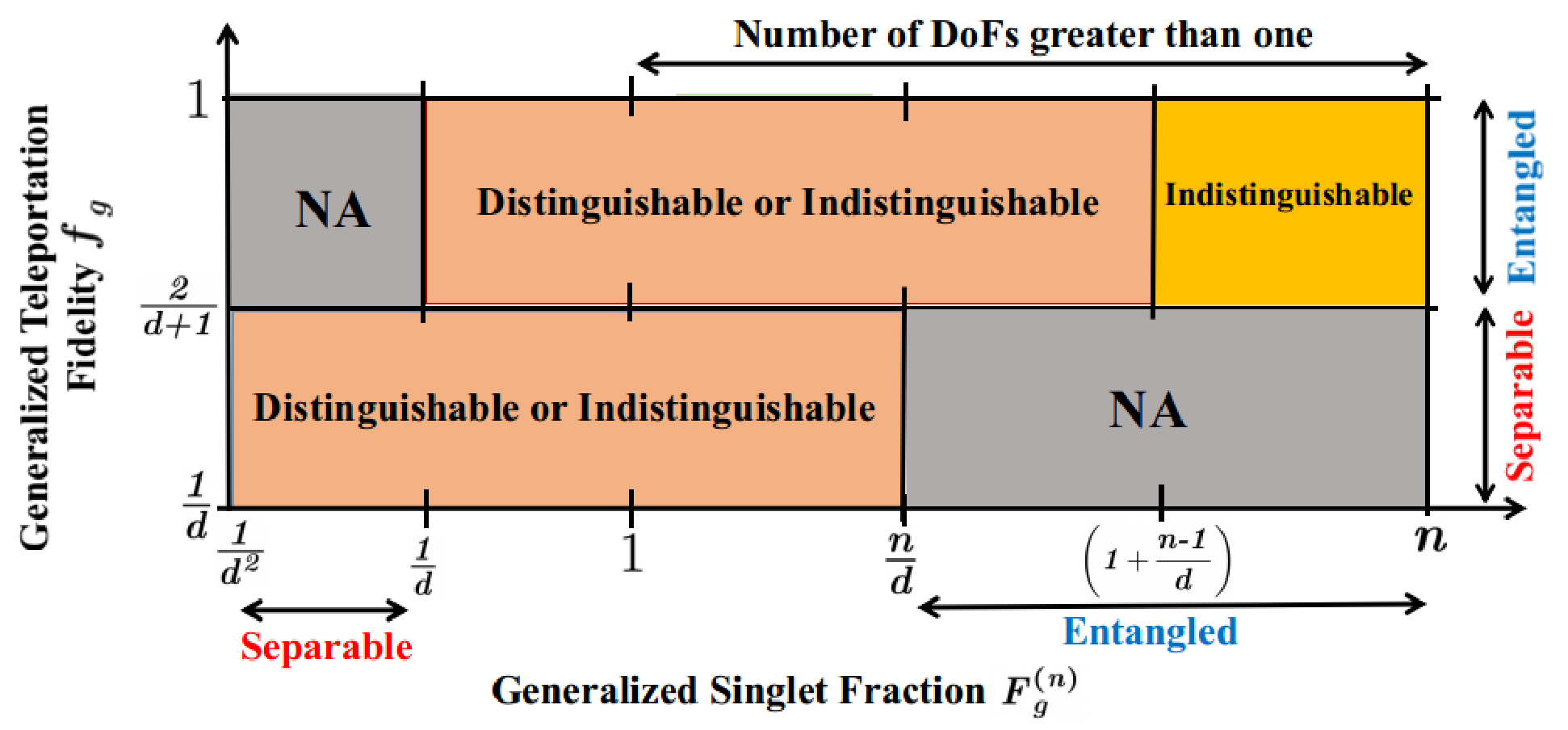} 
\caption{Characterization of the different kinds of states based on their separability, indistinguishability, and number of DoFs $n$ present using generalized singlet fraction $F^{(n)}_{g}$ and generalized teleportation fidelity $f_{g}$.}
\label{TF_SF_rel}
\end{figure*} 

\begin{Theorem}
    The upper bound on the generalized singlet fraction for distinguishable particles is given by
\begin{equation}
\begin{aligned}
F^{(n)}_g 
\leq  \left( 1 + \dfrac{n-1}{d}\right),
\end{aligned}
\end{equation}
for $i, j \in \mathcal{N}$. 
\end{Theorem}
\begin{proof}
Let particle $A$, entangled with $n$ other distinguishable particles $B_1, B_2, \ldots , B_n$, each having a single DoF with dimension $d$, denoted by $\rho^{(1)}_{AB_1B_1 \ldots B_n}$. The singlet fraction between $A$ and $B_j$ for $i \in \mathcal{N}=\lbrace 1, 2, \ldots, n \rbrace$ is
\begin{equation}
F_{AB_j} = \max_{\psi} \braket{\psi\mid \rho^{(1)}_{AB_{j}}\mid \psi},
\end{equation}
where  $\rho^{(1)}_{AB_{j}}=\text{Tr}_{B_{\bar{j}}} (\rho^{(1)}_{AB_1B_1 \ldots B_n}) $ and \\ $B_{\bar{j}}=B_1 B_2 \ldots B_{j-1} B_{j+1} \ldots B_n$. Here, $\ket{\psi}$ varies over all maximally entangled states.
 The monogamy relation with respect to $A$ is given in~\cite{Kay09} as
 \begin{equation} \label{Kay}
 \sum^{n}_{j=1} F_{AB_j} \leq \dfrac{d-1}{d} + \dfrac{1}{n+d-1} \left( \sum^{n}_{j=1} \sqrt{F_{AB_j}} \right)^2. 
\end{equation}  
This relation is also valid if $B_1, B_2, \ldots, B_n$ are the $n$ DoFs of the particle $B$ due to distinguishable scenarios. Our goal is to get the upper bound on $F^{(n)}_g$ for the state $\rho^{(n)}_{AB}$ as defined in Eq.~\eqref{DMdis}.

We take any $n$ numbers of random variables $x_1, x_2 \ldots x_n$ where $ 0 \leq x_j \leq 1$ for all  $j \in \mathcal{N}$. We can write
\begin{equation} \label{Iq}
\begin{aligned}
\left( \sum^{n}_{j=1} \sqrt{x_i} \right)^2 =\sum^{n}_{j=1} x_i + 2 \sum^{n}_{\substack{i, j =1 \\
i > j}} \sqrt{x_ix_j}
\leq  n \left( \sum^{n}_{j=1} x_i \right), 
\end{aligned}
\end{equation}
using the A. M. $\geq$ G.M. inequality  $(x_i+x_j) \geq 2 \sqrt{x_ix_j}$.
From Eq.~\eqref{Iq}, we have
\begin{equation} \label{Fiq}
\left( \sum^{n}_{j=1} \sqrt{F_{A_iB_j}} \right)^2 \leq n \left( \sum^{n}_{j=1} F_{A_iB_j}\right),
\end{equation}
for any $i \in \mathcal{N}$. Now, substituting Eq.~\eqref{Fiq} in Eq.~\eqref{Kay}, we can write
\begin{equation} \label{SumFij}
\sum^{n}_{j=1} F_{A_iB_j}  \leq \left( 1 + \dfrac{n-1}{d}\right). 
\end{equation}
This bound is valid over all the DoFs of $A$, i.e.,
\begin{equation} \label{ij}
\max_{i}  \left\lbrace \sum^{n}_{j=1} F_{A_iB_j} \right\rbrace   \leq \left( 1 + \dfrac{n-1}{d}\right). 
\end{equation}
Similarly, the above bound is valid over all the DoFs of $B$, i.e.,
\begin{equation} \label{ji}
\max_{j}  \left\lbrace \sum^{n}_{i=1} F_{A_iB_j} \right\rbrace   \leq \left( 1 + \dfrac{n-1}{d}\right). 
\end{equation}
From Eq.~\eqref{ij} and Eq.~\eqref{ji}, we have the bound on generalized singlet fraction for distinguishable particles as
\begin{equation}
\begin{aligned}
F^{(n)}_g 
\leq  \left( 1 + \dfrac{n-1}{d}\right),
\end{aligned}
\end{equation}
for $i, j \in \mathcal{N}$. 
\end{proof}

\section{Physical significance of the proposed generalized relation} \label{Phy_sig}

In this section, we characterize different kinds of states based on their separability, indistinguishability, and number of DoFs $n$ present in the state using the knowledge of the generalized singlet fraction $F^{(n)}_{g}$ and the generalized teleportation fidelity $f_{g}$. We find the following cases about any arbitrary two-particle state $\rho$.

\textbf{Case 1:} \textit{Given the values of $d$, $n$, $F^{\left(n \right) }_g$ and $f_g$, can we get the information whether the particles in the state $\rho$ are distinguishable or indistinguishable? }
 
 If we only know the value of $f_g$ then  if $$(f_g=1) \hspace{0.2cm}  \Rightarrow \hspace{0.2cm} \text{$\rho$ is distinguishable},$$    
  because unit fidelity teleportation is not possible for indistinguishable particles~\cite{Das20}, else no conclusion can be drawn.  

 If we know the values of $d$, $n$ and $F^{\left(n \right) }_g$, then  from the Sec.~\ref{UppBound}, we can say 
 
 if $$F^{\left(n \right) }_g > \left( 1+\frac{n-1}{d}\right)   \hspace{0.2cm} \Rightarrow \hspace{0.2cm} \text{$\rho$ is indistinguishable},$$    and no conclusion can be drawn from the converse of the L.H.S.
  
  \begin{figure*}[t!] 
\centering
\includegraphics[width=0.7\textwidth]{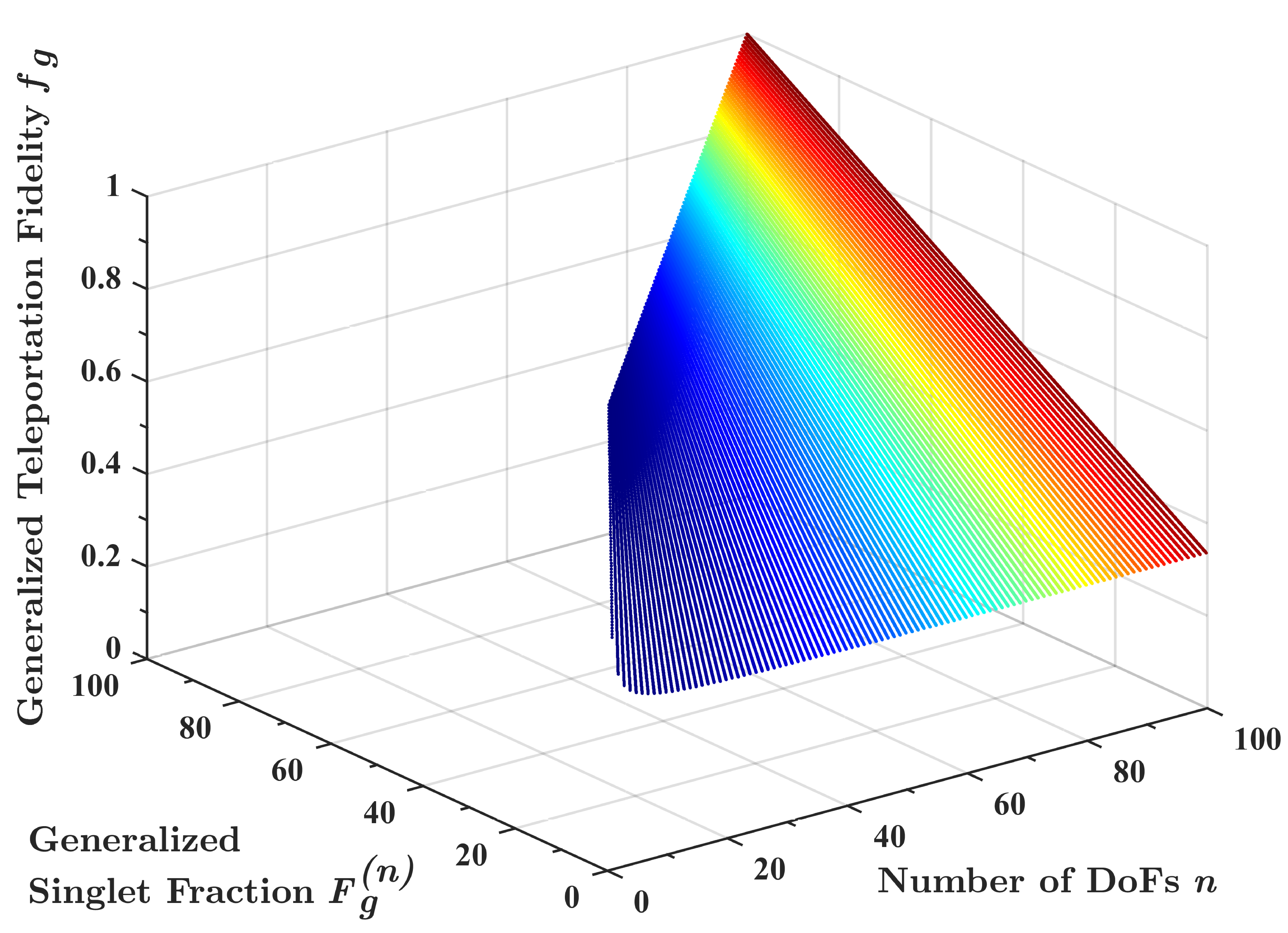} 
\caption{The variation of generalized teleportation fidelity $f_{g}$ and generalized singlet fraction $F^{(n)}_{g}$ with varying the number of DoFs $n$ where the dimension of each DoF is $d=2$.}
\label{TF_SF_graph}
\end{figure*}

\textbf{Case 2:} \textit{Is there any entanglement present in between any two DoFs of $\rho$?}
 
 From the values of $f_g$ or $F^{\left(n \right) }_g$, it can be concluded whether any entanglement structure is present in between any two DoFs of the particles of $\rho$ as follows.
 
 If 
 $$f_g > \left( \frac{2}{d+1}\right)  \hspace{0.1cm}  \Rightarrow \hspace{0.1cm} \text{$\rho$ is entangled}$$ that means at least one entanglement structure is present between any pair of DoFs. Otherwise $\rho$ is separable.
  
  If
  $$ F^{\left(n \right) }_g > \left( \frac{n}{d}  \right) \hspace{0.1cm} \Rightarrow \hspace{0.1cm} \text{$\rho$ is entangled},$$

if
  $$ F^{\left(n \right) }_g \leq  \left( \frac{1}{d}  \right) \hspace{0.1cm} \Rightarrow \hspace{0.1cm} \text{$\rho$ is separable},$$
  otherwise, no conclusion can be drawn.
   
 \textbf{Case 3:}  \textit{How many maximally entangled structures are present in $\rho$?}

 If $\rho$ is entangled then we can find the number of entanglement structures present in it as follows. 

If $f_g=1$,  then the particles are distinguishable, so only one maximally entangled structure is present. 
 
 If $F^{\left(n \right) }_g = n$, the numbers of maximally entangled structures present for any DoF are $n$.

 \textbf{Case 4:} \textit{The number of DoFs present in each particle of $\rho$?}

Here we want to derive the number of DoFs present in each particle of $\rho$
 
 If  $$F^{\left(n \right) }_g >1 \hspace{0.1cm} \Rightarrow \hspace{0.1cm} \text{the number of DoFs}>1$$
  because $F^{(1)}_g \leq 1$ from Eq.~\eqref{f_F_par}. 

 If $$ f_g \leq \frac{2}{d+1} \hspace{0.1cm} \Rightarrow \hspace{0.1cm} \text{no entanglement structure is present,}$$ which implies $$F^{\left(n \right) }_g=\left( \frac{n}{d} \right) \hspace{0.2cm} \text{or}, \hspace{0.2cm} n=\ceil{ dF^{\left(n \right) }_g }.$$ 

All the above cases are pictorially represented in Fig.~\ref{TF_SF_rel}. 

For $d=2$, the relation between $n$, $F^{\left(n \right) }_g$, and $f_g$ are plotted in Fig.~\ref{TF_SF_graph} for $1 \leq n \leq 100$ using Eq.~\eqref{fandF}. 
 \label{QPQ_app}

\section{Illustration of the generalized singlet fraction using optical circuits} \label{Op_Ckt}
If two particles each having a single DoF are maximally entangled, then the value of the generalized singlet fraction is one. However, the converse is not necessarily true which we illustrate by proposing an optical circuit using two distinguishable particles each having two DoFs, polarization, and orbital angular momentum.

\begin{figure*}[t] 
\centering
\includegraphics[width=0.8\textwidth]{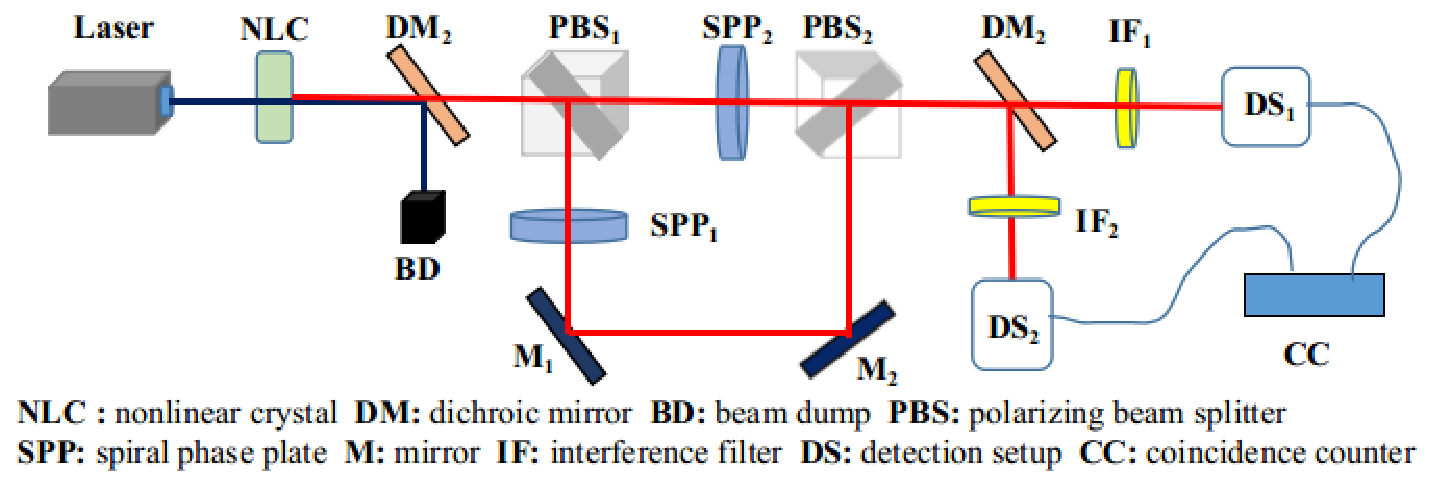} 
\caption{Optical set up to the generation of entangled state between polarization and orbital angular momentum DoF for distinguishable particles.
NLC: nonlinear crystal, DM: dichroic mirror, BD: beam dump, PBS: polarizing beam splitters, SPP: spiral phase plates, M: mirrors, IF: interference filters, DS: detection setup and C.C.: coincidence counter. The detection setups could be made up of spatial light modulators and single-mode fibers coupled to avalanche photodetectors if measuring in orbital angular momentum or half-wave plates and PBS if measuring in polarization.
}
\label{Dis_HHES_ckt}
\end{figure*}

In Fig.~\ref{Dis_HHES_ckt}, a nonlinear  crystal ($\beta$ barium borate, periodically poled potassium triphosphate etc.) is pumped by a pulsed laser from which a pump photon is absorbed, and two photons are generated governed by the phase-matching conditions:  
$
\omega_p=\omega_s + \omega_i, \hspace{0.2cm} \textbf{k}_p = \textbf{k}_s+\textbf{k}_i
$
 where $\hbar\omega_j$ is the energy and $\hbar\textbf{k}_j$ is the momentum of the \emph{j}-th photon. The indices \emph{p}, \emph{s} and \emph{i} stand for pump, signal, and idler respectively. The signal and idler photon pair produced in this method under non-degenerate type-II spontaneous parametric down-conversion are correlated in polarization, and are represented as 

 \begin{equation}
 \vert \Psi_{i} \rangle = \cos\theta \vert H\rangle_{\omega_s}\vert V \rangle_{\omega_i} + e^{i\phi}\sin\theta \vert V\rangle_{\omega_s} \vert H \rangle_{\omega_i},
 \end{equation}
 
\noindent where $\ket{H}$ and $\ket{V}$ denote horizontal and vertical polarization, $\theta$ controls the normalization factor, and $\phi$ is a phase term that arises from birefringence in the nonlinear crystal. Since $\omega_s\neq\omega_i$ (under non-degenerate phase matching) this leads to distinguishability between the signal and idler photons. Let this state be now incident on a broadband polarizing beam splitter which allows $\vert H \rangle$ photons to pass through to the transmitted path, while $\vert V \rangle$ photons are directed to the reflected path.
Each of the paths contains a spiral phase plate of equal order $l$ which can take any value from $-1$ to $+1$ through $0$. Spiral phase plates are optical devices with continuously varying thicknesses. Photons pick up orbital angular momentum of the corresponding order $l$ on passing through them.
The two paths are again combined at another broadband polarizing beam splitter resulting in the state

\begin{equation} \label{DIsHHES} 
\vert \Psi_{f} \rangle = \cos\theta \vert H, +l\rangle_{\omega_s}\vert V, -l \rangle_{\omega_i} + e^{i\phi}\sin\theta \vert V, -l\rangle_{\omega_s} \vert H, +l \rangle_{\omega_i}.
\end{equation}

The handedness of \emph{l} is sensitive to reflections and changes by $\exp(i\pi)$ under each reflection. The difference in the handedness of \emph{l} arises in the above equation due to an unequal number of reflections between the transmitted and reflected arms.
Now we calculate the generalized singlet fraction of our proposed state. 
Tracing out both the orbital angular momentum DoFs from the density matrix $\rho_f=\ket{\Psi_{f}}\bra{\Psi_{f}} $, we get
\medmuskip=0mu
\thinmuskip=0mu
\thickmuskip=0mu
\begin{small}
\begin{equation}\label{rho21wswi}
\begin{aligned}
\rho_{f_{p}} = & \cos^{2}\theta  \vert H\rangle_{\omega_s}\vert V\rangle_{\omega_i} \bra{ H}_{\omega_s}\bra{V }_{\omega_i} + \sin^{2}\theta \ket{ V}_{\omega_s} \ket{ H }_{\omega_i} \bra
{ V }_{\omega_s} \bra{ H }_{\omega_i}.
\end{aligned}
\end{equation}
\end{small}
\medmuskip=2mu
\thinmuskip=2mu
\thickmuskip=2mu
\noindent The singlet fraction for Eq.~\eqref{rho21wswi} is 
\begin{equation}
\max_{\psi} \braket{\psi|\rho_{f_{p}}|\psi} = \frac{1}{2} (\cos^{2} \theta + \sin^{2} \theta ) =1/2.
\end{equation}
 Similarly, for other entanglement connections, the value is the same, i.e., $1/2$. So, the generalized singlet fraction of the state $\vert \Psi_{f} \rangle$ using Eq.~\eqref{SFGen} is $F^{(2)}_g =1$.
A similar conclusion can be drawn for two indistinguishable particles as proposed in~\cite{Bhatti15}. Note that, our proposed state in Eq.~\eqref{DIsHHES}is different from~\cite{Bhatti15} as we use non-degenerate phase-matching to make use of distinguishable photons.
Thus, the generalized singlet fraction alone might not be the best quantifier for the presence of maximal entanglement.

 Another special state is a hyper-hybrid entangled state~\cite{HHNL} with two particles each having two DoFs. The maximum value of $F^{\left(2 \right) }_g=2 $ for this state~\cite[Eq. (4)]{HHNL} is achieved as each DoF of one particle is maximally entangled with all the other DoFs of other particles as shown in~\cite{Paul21}.

\section{A cryptographic resource reduction using generalized singlet fraction} \label{Cryp_app}

Security and efficiency are the two major criteria of a cryptographic protocol.
 If two cryptosystems with different resource requirements provide the same level of security, the one with fewer resources becomes the natural choice. Here, we present a generic scheme to reduce the number of particles used in device-independent~\cite{Acin07} quantum private query~\cite{QPQ07,DIQPQ} without affecting the security.
In quantum cryptography, DI tests~\cite{Acin07} are required to establish a secure key between two parties when the devices are not trustworthy.
In a quantum private query, a user queries a database about some specific entry and gets back only the corresponding data without revealing the query to the server. To achieve this, a secure key must be established between the database owner and the user such that the database owner knows the full key but the user knows a fraction of that key. In device-independent quantum private query~\cite{DIQPQ} using Clauser-Horne-Shimony-Holt test~\cite{CHSH}, the maximum success probability is $\cos^{2}\frac{\pi}{8} \approx 85\%$ using two qubits. 
With an ancilla particle, this success probability can be increased asymptotically to unity using quantum pseudo-telepathy game~\cite{Brassard05,Jyoti18}. Details of these protocols can be found in Appendix~\ref{DIQPQapp}.

The cost of adding an ancilla particle can be bypassed by using an additional DoF and creating multi-DoF entanglement where the success probability remains the same but the generalized singlet fraction changes. 
In the key generation phase of quantum private query protocol as given in~\cite{Yang14}, the source shares an entangled pair of particles with Bob and Alice are in the specific form
\begin{equation} \label{QPQstatemain}
\ket{\psi}_{BA}=\frac{1}{\sqrt{2}}\left( \ket{0}_{B}\ket{\phi_{0}}_{A} + \ket{1}_{B}\ket{\phi_{1}}_{A}\right)
\end{equation}
where
\begin{eqnarray*}
\ket{\phi_{0}}_{A} = \text{cos} \left( \frac{\theta}{2}\right) \ket{0} + \text{sin} \left( \frac{\theta}{2}\right) \ket{1}, \\
\ket{\phi_{1}}_{A} = \text{cos} \left( \frac{\theta}{2}\right) \ket{0} - \text{sin}\left( \frac{\theta}{2}\right) \ket{1},
\end{eqnarray*}
and $0 < \theta < \frac{\pi}{2}$.
With the help of an ancilla qubit $X$, the state
in Eq.~\eqref{QPQstatemain} can be transformed into the following state
 \begin{widetext}
 \begin{equation} \label{pseudo_statemain}
 \ket{\psi}_{BAX} = \frac{1}{\sqrt{2}} \left( \text{cos}\frac{\theta}{2} \ket{000}_{BAX}  + \text{sin}\frac{\theta}{2} \ket{010}_{BAX} + \text{cos}\frac{\theta}{2} \ket{111}_{BAX} - \text{sin}\frac{\theta}{2} \ket{100}_{BAX} \right). 
 \end{equation}
 \end{widetext}
 \begin{figure*}[t!] 
\centering
\includegraphics[width=0.7\textwidth]{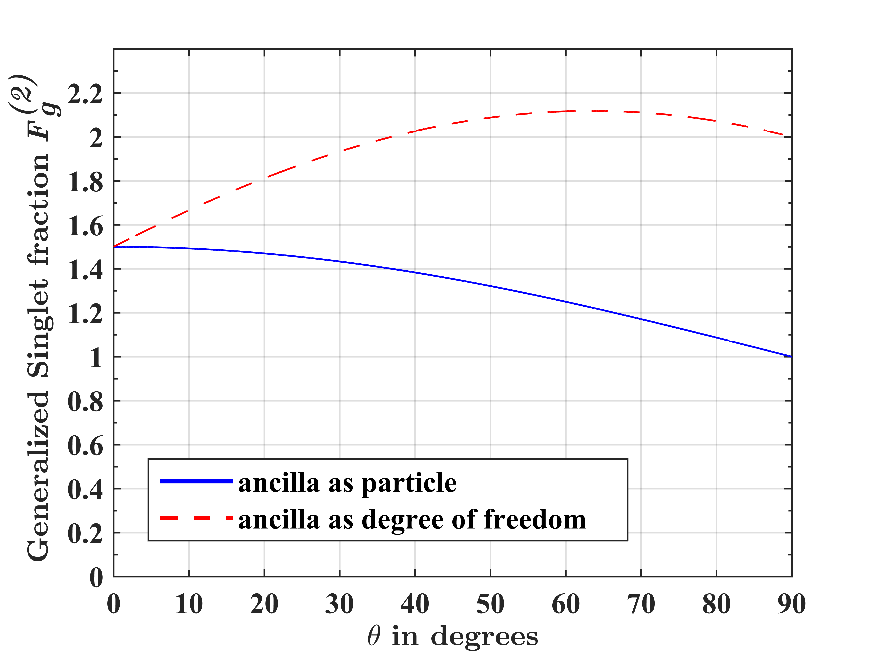} 
\caption{The variation of generalized singlet fraction $F^{(n)}_{g}$ for the quantum pseudo-telepathy test using ancilla as a particle and using ancilla as degrees of freedom with varying $\theta$ in degrees.}
\label{Appli}
\end{figure*}
 Now, instead of choosing the ancilla as another particle, if another DoF of $A$ is used, we can reduce the number of particles used in the quantum pseudo-telepathy test. Then the state in Eq.~\eqref{pseudo_statemain} can be written as
\begin{widetext}
 \begin{equation} \label{pseudo_state_DoFmain}
 \ket{\psi}_{BA_1A_2} = \frac{1}{\sqrt{2}} \left( \text{cos}\frac{\theta}{2} \ket{000}_{BA_1A_2}  + \text{sin}\frac{\theta}{2} \ket{010}_{BA_1A_2} + \text{cos}\frac{\theta}{2} \ket{111}_{BA_1A_2} - \text{sin}\frac{\theta}{2} \ket{100}_{BA_1A_2} \right). 
 \end{equation}
\end{widetext} 
 Assuming each particle have two DoFs, i.e., $n=2$,  the generalized singlet fraction for the state in Eq.~\eqref{pseudo_statemain} is
 \begin{widetext}
 \begin{equation}
 \begin{aligned}
 F^{(2)}_{g} (\ket{\psi}\bra{\psi}_{BAX})&=\max \left\{  \max_{i} \left\lbrace  \max_{\psi_{a_{i}b_{{j}}}} \left\lbrace  \sum^{2}_{j=1} \braket{\psi_{a_{i}b_{j}} \mid \rho_{a_{i}b_{j}} |\psi_{a_{i}b_{j}}}  \right\rbrace \right\rbrace ,   \max_{j} \left\lbrace  \max_{\psi_{a_{i}b_{{j}}}} \left\lbrace  \sum^{2}_{i=1} \braket{\psi_{a_{i}b_{j}} \mid \rho_{a_{i}b_{j}} |\psi_{a_{i}b_{j}}}  \right\rbrace  \right\rbrace  \right\}\\
 &=\frac{1}{2}+\cos^{2} \frac{\theta}{2}
 \end{aligned}
 \end{equation}
 \end{widetext} 
where $\rho_{a_{i}b_{j}}:= \text{Tr}_{a_{\bar{i}}{b_{\bar{j}}}} ( \rho^{(2)}_{AB} )$ and $\rho^{(2)}_{AB}=\text{Tr}_{X}\left( \ket{\psi}_{BAX} \bra{\psi}_{BAX}\right) $. 

Similarly, if we calculate the generalized single fraction for the state in Eq.~\eqref{pseudo_state_DoFmain}, we get 
\begin{equation}
F^{(2)}_{g} (\ket{\psi}\bra{\psi}_{BA_1A_2})=\frac{1}{2}+\cos^{2} \frac{\theta}{2} + 2\cos \frac{\theta}{2}\sin \frac{\theta}{2}.
\end{equation}

Thus, for the system using an additional ancilla particle~\cite{Jyoti18}, we show that the generalized singlet fraction is $(\frac{1}{2}+\cos^{2} \frac{\theta}{2})$ where $0 < \theta < \pi/2$. But if an additional DoF is used instead, this value becomes $(\frac{1}{2}+\cos^{2} \frac{\theta}{2} + 2\cos \frac{\theta}{2}\sin \frac{\theta}{2})$ ). 

Now one can see that although the states in Eq.~\eqref{pseudo_statemain} and   Eq.~\eqref{pseudo_state_DoFmain} are physically different in terms of the number of particles and the number of DoFs present, mathematically they have the same form. Thus it is challenging to characterize and analyse these states. Using our generalized singlet fraction, we can easily differentiate and characterize these states. The values of the generalized singlet fraction for the states for the quantum pseudo-telepathy test using ancilla as a particle in Eq.~\eqref{pseudo_statemain} and using ancilla as DoFs Eq.~\eqref{pseudo_state_DoFmain} are shown in Fig.~\ref{Appli}. It can be seen that as $\theta$ increases, the values of generalized singlet fraction decreases for Eq.~\eqref{pseudo_statemain} but increases for  Eq.~\eqref{pseudo_state_DoFmain}. This clearly differentiate these two state mathematically.

 Thus, not only for this application, there are many ancilla-assisted quantum processing protocols like quantum process tomography~\cite{Altepeter03}, entanglement stabilization~\cite{Andersen19}, quantum secret sharing~\cite{Mitra15}, quantum error correction~\cite{Criger12}, quantum measurement~\cite{ Brida12}, weak value amplification~\cite{Pang14}, quantum channel discrimination~\cite{Bae19}, coherent-state superposition generation~\cite{Takahashi08}, etc., the resource can be reduced by using additional DoF as ancilla instead of a particle. Further, the states can also be differentiated using our generalized singlet fraction.

\section{Discussion} \label{Diss}

We offer a generalization of the original relation between teleportation fidelity and singlet fraction for distinguishable particles with a single DoF to physical settings where either distinguishable or indistinguishable particles with single or multiple DoFs are involved. 

The earlier work in~\cite{Horodecki99} considered only a single DoF of distinguishable particles.
However, our work is not directly comparable with~\cite{Horodecki99} because the notion of distinguishable particles is well-established and based on the tensor product of the corresponding Hilbert space formalism. But, for indistinguishable particles, the notion of entanglement is represented by mainly two methods, particle-based method~\cite{Li01,You01,John01,Zanardi02,Ghirardhi02,Wiseman03,Ghirardi04} and mode-based method~\cite{Vedral03,Barnum04,Zanardi04}. The other methods are variants of these two methods~\cite{Benatti14,Benatti17,Benatti20}. 

In these two methods, the notion of entanglement sometimes provides independent separated particles as the entangled ones~\cite{Ghirardi04,Tichy11,Braun18}. Recently, Lo Franco \textit{et al.}~\cite{LFC16,LFC18} have proposed a particle-based method that overcomes these drawbacks. However, their method only applies to indistinguishable particles with a single DoF. Later, Paul \textit{at al.}~\cite{Paul21} proposed a generalized version with is applicable for distinguishable and indistinguishable particles with single or multiple DoFs. We have used the method of Paul \textit{et al.}~\cite{Paul21} to derive our relations which is the most generalized particle-based representation till now.

One may question that our proposed generalized relation is not unique since it is based on the particle-based version and it may be different in the mode-based version. This confusion is clarified in~\cite{Lourenco19} which states that the particle-based approach of Lo Franco \textit{et al.} is equivalent to a modified version of the mode-based approach. Thus if one follows their approach, the same generalized relation can be found. So, our generalized relation is not based on some specific approach and can be treated as a universal one.

 One may wonder if using a particle as an ancilla is the same as using an additional DOF of the particle as an ancilla. Mathematically they are the same. However, for practical scenarios, if
we use particle as an ancilla then it increases the resource. Thus we propose to use additional DOF of the particle as ancilla which reduces the
resource. Further, we analyze how using ancilla as a particle and ancilla as a DoF can change the value of the generalized
singlet fraction for the quantum pseudo-telepathy test. This can be used to analyze the states which are mathematically same but physically different in terms of the number of particles and the number of DoFs.

\section*{Acknowledgements.} SD and GP would like to thank Dr. Ritabrata Sengupta for various discussions. S. Das would like to acknowledge the support received by Quantum Delta NL KAT-2 project for the postdoctoral fellowship during part of the work. AB would like to thank the Physical Research Laboratory, Ahmedabad for the postdoctoral fellowship during which this work was done.\\

\appendix
\section*{APPENDIX}
\section{Representation for two distinguishable and indistinguishable particles} \label{Tr_out_rule}
In this section, we will represent the general state, density matrix, and the degree of freedom (DoF) trace-out rule for two distinguishable and indistinguishable particles using the formalism of~\cite{LFC16,LFC18,Paul21}.

\begin{definition}
Let two distinguishable particles $A$ and $B$ each having $n$ DoFs. The $i$-th and the $j$-th DoF of $A$ and $B$ are represented by $a_{i}$ and $b_{j}$  respectively where $i, j \in \mathcal{N} :=\lbrace 1, 2, \ldots , n \rbrace$.  Suppose each DoF is $d$-dimensional whose eigenvalues are denoted by
$\mathbb{D}_{k}:=\lbrace D_{k_{1}}, D_{k_{2}}, \ldots, D_{k_{d}} \rbrace$ where $k \in \mathcal{N}$. The general state of $A$ and $B$ is defined as 
\begin{equation} \label{Statedis}
\begin{aligned}
\ket{\Psi^{(n)}}_{AB}= \sum_{a_{1}, a_{2}, \ldots , a_{n},  b_{1}, b_{2}, \ldots, b_{n}}\kappa^{ a_{1} a_{2} \ldots a_{n}}_{  b_{1} b_{2} \ldots b_{n}} \ket{ a_{1} a_{2} \ldots a_{n}} \otimes \ket{  b_{1} b_{2} \ldots b_{n} },
\end{aligned}
\end{equation}
where $a_i \in \mathbb{D}_{i}$, $b_j \in \mathbb{D}_{j}$, and $i,j \in \mathcal{N}$.

The general density matrix can be represented as 
\begin{equation} \label{DMdisapp}
\begin{aligned}
\rho^{(n)}_{AB}= \sum_{\substack{ a_{1}, a_{2}, \ldots, a_{n}, u_{1}, u_{2}, \ldots, u_{n}  \\   b_{1}, b_{2}, \ldots, b_{n}, v_{1}, v_{2}, \ldots, v_{n}}} \kappa^{ a_{1} a_{2} \ldots a_{n} u_{1} u_{2} \ldots u_{n} }_{  b_{1} b_{2} \ldots b_{n}  v_{1} v_{2} \ldots v_{n}}  \ket{ a_{1} a_{2} \ldots a_{n}} \ket{  b_{1} b_{2} \ldots b_{n}} \otimes \bra{ u_{1} u_{2} \ldots u_{n}}\bra{  v_{1} v_{2} \ldots v_{n}},
\end{aligned}
\end{equation}
where $a_i, u_i \in \mathbb{D}_{i}$, $b_j, v_j \in \mathbb{D}_{j}$, and $i,j \in \mathcal{N}$. If $\rho^{(n)}_{AB} = \ket{\Psi^{(n)}}_{AB}\bra{\Psi^{(n)}}_{AB}$ then $\kappa^{ a_{1} a_{2} \ldots a_{n} u_{1} u_{2} \ldots u_{n} }_{  b_{1} b_{2} \ldots b_{n}  v_{1} v_{2} \ldots v_{n}}$ can be factorized as $\kappa^{ a_{1} a_{2} \ldots a_{n}}_{  b_{1} b_{2} \ldots b_{n}} \kappa^{ u_{1} u_{2} \ldots u_{n}*}_{  v_{1} v_{2} \ldots v_{n}}$ where $*$ denotes complex conjugate.
\end{definition}

 If we want to trace-out the $i$-th DoF of particle $A$, then from Eq.~\eqref{DMdisapp}, the reduced density matrix can be written as
  \begin{widetext}
\begin{equation} \label{DisDoFTraceout}
\begin{aligned}
\rho_{a_{\bar{i}}} \equiv & \text{Tr}_{a_{i}} \left( \rho^{(n)}_{AB} \right) := \sum_{ a_{i}, a_{\bar{i}}, u_{i} u_{\bar{i}},  b_{1}, b_{2}, \ldots, b_{n} , v_{1}, v_{2}, \ldots, v_{n}} \kappa^{ a_{\bar{i}} u_{\bar{i}}}_{ b_{1} b_{2} \ldots b_{n} v_{1} v_{2} \ldots v_{n}} \ket{ a_{\bar{i}}} \ket{   b_{1} b_{2} \ldots b_{n} }  \bra{u_{\bar{i}}} \bra{ v_{1} v_{2} \ldots v_{n}}
 \left\lbrace \braket{a_{i}|c_{i}} \right\rbrace ,
\end{aligned}
\end{equation}
 \end{widetext} 
where $a_{\bar{i}}=a_{1}a_{2} \ldots a_{i-1}a_{i+1} \ldots a_{n}$ and similar meaning for $u_{\bar{i}}$. One can show that when the DoF trace-out rule in Eq.~\eqref{DisDoFTraceout} is applied to the same particle for $n$ times, it becomes equivalent to our familiar particle trace-out rule~\cite[Eq. 2.178]{Nielsenbook}.

 The DoF trace-out rule when each indistinguishable particle has $n$ DoFs and a spatial label as proposed by Paul \textit{et al}~\cite{Paul21} can be explained as follows. 

 \begin{definition}
Lets us consider two indistinguishable particles each having $n$ DoFs, having $p$ localized regions denoted by $ \mathcal{S} = \left\lbrace s_{1}, s_{2}, \cdots , s_{p} \right \rbrace $. The $i$-th and $j$-th DoF are represented by $a_{i}$ and $b_{j}$ respectively, where $i, j \in \mathcal{N} =\lbrace 1, 2, \ldots , n \rbrace$.  Suppose each DoF is $d$-dimensional whose eigenvalues are denoted by
$\mathbb{D}_{k}:=\lbrace D_{k_{1}}, D_{k_{2}}, \ldots, D_{k_{d}} \rbrace$ where $k \in \mathcal{N}$. The general state of two particles can be represented by 
\begin{widetext}
\begin{equation} \label{State}
\begin{aligned}
\ket{\Psi^{(n)}}= \sum_{\alpha,  a_{1}, a_{2}, \ldots , a_{n}, \beta, b_{1}, b_{2}, \ldots, b_{n}}\eta^{\nu} \kappa^{\alpha a_{1} a_{2} \ldots a_{n}}_{\beta  b_{1} b_{2} \ldots b_{n}} \ket{\alpha a_{1} a_{2} \ldots a_{n}, \beta  b_{1} b_{2} \ldots b_{n} }.
\end{aligned}
\end{equation}
\end{widetext}
Here, $  \alpha , \beta$ ranges over $ \mathcal{S}  $. Each of  $a_{i}$ and $b_{j}$ ranges over $ \mathcal{D}_{i}$ and $ \mathcal{D}_{j}$ respectively where $i, j \in \mathcal{N} $. Here, $\nu$ denotes the sum of the parity of the cyclic permutations for all the DoFs where $\eta=+1$ for bosons and $\eta=-1$ for fermions. If  
$\lbrace (\alpha=\beta ) \wedge ( a_{k}=b_{k} ) \rbrace$ 
 for any  $k \in \mathcal{N}$, then $\eta=0$ for fermions due to the Pauli exclusion principle~\cite{Pauli25}.

 The general density matrix of two indistinguishable particles can be described as
 \medmuskip=0mu
\thinmuskip=0mu
\thickmuskip=0mu
\begin{widetext}
\begin{equation} \label{DMapp}
\begin{aligned}
\rho^{(n)}= \sum_{\substack{\alpha,   a_{1}, a_{2}, \ldots, a_{n}, \gamma,  u_{1}, u_{2}, \ldots, u_{n} \\  \beta, b_{1}, b_{2}, \ldots, b_{n}, \delta,  v_{1}, v_{2}, \ldots, v_{n}}} \eta^{\mu}\kappa^{\alpha a_{1} a_{2} \ldots a_{n} \gamma u_{1} u_{2} \ldots u_{n}}_{\beta  b_{1} b_{2} \ldots b_{n}\delta  v_{1} v_{2} \ldots v_{n}}  \ket{\alpha a_{1} a_{2} \ldots a_{n}, \beta  b_{1} b_{2} \ldots b_{n}} \bra{\gamma u_{1} u_{2} \ldots u_{n}, \delta  v_{1} v_{2} \ldots v_{n}},
\end{aligned}
\end{equation}
\end{widetext}
 \medmuskip=2mu
\thinmuskip=2mu
\thickmuskip=2mu
where $ \alpha, \beta ,\gamma , \delta$ span $ \mathcal{S} $. Also $a_{i}, u_{i}$ span $ \mathcal{D}_{i}$ and  $ b_{j}, v_{j}$ span $ \mathcal{D}_{j}$  with $i, j \in  \mathcal{N}$. Here, $\mu=\nu + \bar{\nu}$ denotes the sum of the parity of the cyclic permutations for all the DoFs where $\eta=+1$ for bosons and $\eta=-1$ for fermions. Here $\bar{\nu}$ comes due to the density matrix. If  
$\lbrace (\alpha=\beta ) \wedge ( a_{k}=b_{k} ) \rbrace$  or $ \lbrace (\gamma=\delta )   \wedge ( u_k=v_k ) ) \rbrace $
 for any  $k \in \mathcal{N}$, then $\eta=0$ for fermions due to the Pauli exclusion principle~\cite{Pauli25}. 
 \end{definition}
 
\section{DoF trace-out rule for indistinguishable particles} \label{Parti_DoF_Tr}
 
Next, we define the DoF trace-out rule for indistinguishable particles from the general density matrix of two particles as defined in Eq.~\eqref{DMapp}. Suppose we want to trace-out the $i$-th DoF of location $s_x \in \mathcal{S}$. Then the DoF reduced density matrix is
 \medmuskip=0mu
\thinmuskip=0mu
\thickmuskip=0mu
 \begin{widetext}
\begin{equation} \label{DOFTrRule}
\begin{aligned}
\rho_{s_{x_{\bar{i}}}} \equiv  \text{Tr}_{s_{x_{i}}} \left(  \rho^{(n)} \right) \equiv & \sum_{m_{i}} \braket{ s_x m_i \mid \rho^{(n)} \mid s_x m_i } \\
:=& \eta^{\mu}\sum_{m_{i}} \bigg \lbrace \sum_{\substack{ \alpha, \beta, a_{i}, a_{\bar{i}},  b_{1}, b_{2}, \ldots, b_{n}, \\ \gamma, \delta, u_{i}, u_{\bar{i}}, v_{1}, v_{2}, \ldots, v_{n}} } \kappa^{\alpha a_{i} a_{\bar{i}} \gamma u_{i} u_{\bar{i}}}_{\beta b_{1} b_{2} \ldots b_{n} \delta v_{1} v_{2} \ldots v_{n}} \braket{s_x m_i \mid \alpha a_i} \braket{\gamma u_i  \mid s_x m_i}  \ket{\alpha a_{\bar{i}}, \beta  b_{1} b_{2} \ldots b_{n}}\bra{\gamma u_{\bar{i}}, \delta  v_{1} v_{2} \ldots v_{n}}  \\
& +  \eta \sum_{\substack{ \alpha, \beta,  a_{1}, a_{2}, \ldots, a_{n}, b_{i}, b_{\bar{i}}, \\ \gamma, \delta, u_{i}, u_{\bar{i}}, v_{1}, v_{2}, \ldots, v_{n}} } \kappa^{\alpha a_{1} a_{2} \ldots a_{n} \gamma u_{i} u_{\bar{i}}}_{\beta b_{i} b_{\bar{i}} \delta v_{1} v_{2} \ldots v_{n}}  \braket{s_x m_i \mid \beta b_i }  \braket{\gamma u_i  \mid s_x m_i}  \ket{\alpha a_{1} a_{2} \ldots a_{n}, \beta  b_{\bar{i}}} \bra{\gamma u_{\bar{i}}, \delta  v_{1} v_{2} \ldots v_{n}} 
\\
&+ \eta  \sum_{\substack{ \alpha, \beta, a_{i}, a_{\bar{i}},  b_{1}, b_{2}, \ldots, b_{n}, \\ \gamma, \delta, u_{1}, u_{2}, \ldots, u_{n} , v_{i}, v_{\bar{i}}} } \kappa^{\alpha a_{i} a_{\bar{i}}\gamma u_{1} u_{2} \ldots u_{n}}_{\beta b_{1} b_{2} \ldots b_{n}\delta v_{i} v_{\bar{i}}}  \braket{s_x m_i \mid \alpha a_i}  \braket{\delta v_i  \mid s_x m_i}  \ket{\alpha a_{\bar{i}}, \beta  b_{1} b_{2} \ldots b_{n}}\bra{\gamma u_{1} u_{2} \ldots u_{n}, \delta  v_{\bar{i}}}   \\
 & + \sum_{\substack{ \alpha, \beta,  a_{1}, a_{2}, \ldots, a_{n}, b_{i}, b_{\bar{i}}, \\ \gamma, \delta, u_{1}, u_{2}, \ldots, u_{n} , v_{i}, v_{\bar{i}}} } \kappa^{\alpha a_{1} a_{2} \ldots a_{n}\gamma u_{1} u_{2} \ldots u_{n}}_{\beta b_{i} b_{\bar{i}}\delta v_{i} v_{\bar{i}}}  \braket{s_x m_i \mid \beta b_i } \braket{\delta v_i  \mid s_x m_i}   \ket{\alpha a_{1} a_{2} \ldots a_{n}, \beta  b_{\bar{i}}} \bra{\gamma u_{1} u_{2} \ldots u_{n}, \delta  v_{\bar{i}}} \bigg \rbrace ,
\end{aligned}
\end{equation}
 \end{widetext}
 \medmuskip=2mu
\thinmuskip=2mu
\thickmuskip=2mu
where $a_{\bar{i}} := a_{1}a_{2} \ldots a_{i-1}a_{i+1} \ldots a_{n}$ and similar meaning for  $b_{\bar{i}}$, $u_{\bar{i}}$ and $v_{\bar{i}}$. 

\section{Proof of relation between the teleportation fidelity and the singlet fraction for any state} \label{HorodeckiProof}
In~\cite{Horodecki99}, Horodecki \textit{et al.} showed a relation~\cite[Eq. (16)]{Horodecki99} between the teleportation fidelity $(f)$ and the singlet fraction $(F)$ for distinguishable particles as shown in Eq.~\eqref{f_F_par}. They also proved its validity for the noisy singlet states or isotropic states or generalized Werner states~\cite{Werner89}  of the form 
\begin{equation} \label{rho_NS_dis}
\rho_{p}= p P^{+} + (1-p) \dfrac{\mathbb{I} \otimes \mathbb{I}}{d^2}, \hspace{1cm} 0 \leq p \leq 1
\end{equation}
where each particle has a single DoF and
\begin{equation} \label{P+}
P^{+} = \ket{\psi^{+}} \bra{\psi^{+}},  \hspace{0.2cm} \ket{\psi^{+}}=\frac{1}{\sqrt{d}} \sum^{d-1}_{i=0} \ket{i} \ket{i},
\end{equation}
and $d$ is the dimension of the maximally entangled state $\ket{\psi^{+}}$ using distinguishable particles.

 Then they have proved that their relation is true for any state $\rho$~\cite{Horodecki99}. We summarise their proof in our own words for easy reference.
 
 First, they proved that a set of channels $\Lambda$ is isomorphic to the set of density matrices $\rho$. Next, they showed that any state $\rho$ under $U \otimes U^{*}$ twirling operation~\cite{Werner89} produces a noisy singlet $\rho_{p}$ with the same fidelity $F$ as the original state $\rho$ where the $*$ denotes complex conjugation. After that, they proved a depolarizing channel $\Lambda^{\text{dep}}$ is equivalent to noisy singlets denoted by $\rho_{p}$. Further, they proved any channel $\Lambda$ subjected to a twirling process denoted by $\mathcal{T}$ becomes a depolarizing channel. 
 Now using the equivalence between depolarizing channel and noisy singlets, they wrote the relation between teleportation fidelity and singlet fraction in Eq.~\eqref{f_F_par} for any depolarizing channel $\Lambda^{\text{dep}}$. Further, they proved that the teleportation fidelity $f$ and the singlet fraction $F$ are invariant of the twirling process $\mathcal{T}$. Thus the relation in Eq.~\eqref{f_F_par} is also true for any channel $\Lambda$. Finally, 
 from the isomorphism between channel and state, they showed that Eq.~\eqref{f_F_par} is true for any state $\rho$.
 
\begin{enumerate}
\item A set of channels $\Lambda$ is isomorphic to the set of density matrices $\rho$ represented as
\begin{equation} \label{Lab_rho}
\rho \equiv \Lambda.
\end{equation}
They have proved this in the following way:

\textit{A. Channel $\rightarrow$ State} For a given channel~\cite{Nielsenbook} denoted by $\Lambda$, it is possible to ascribe it to a state $\rho_{\Lambda}$ such that
\begin{equation} \label{Channel2State}
\rho_{\Lambda} = (I \otimes \Lambda) P^{+},
\end{equation}
where $P^{+}$ is as defined in Eq.~\eqref{P+}. 

\textit{B. State $\rightarrow$ Channel } For any given state $\rho$ there is always a unique channel $\Lambda_{\rho}$ such that it satisfies
\begin{equation} \label{State2Channel}
\rho = (I \otimes \Lambda_{\rho}) P^{+}.
\end{equation}

 \item Any state $\rho$ under $U \otimes U^{*}$ twirling operation~\cite{Werner89} produces a noisy singlet $\rho_{p}$ with the same fidelity $F$ as the original state $\rho$ where the $*$ denotes complex conjugation.  The key idea behind the proof is to show that the singlet states as shown in Eq~\eqref{P+} are invariant under $U \otimes U^{*}$ twirling operations. All the other states become random noise under $U \otimes U^{*}$ twirling operations. Thus to prove that one has to show that noisy singlets are invariant under $U \otimes U^{*}$ twirling operations. Note that, this twirling operation does not create a new singlet, it just preserves the singlet while averaging out all other terms~\cite{Anwar05}. 

\item A depolarizing channel~\cite{HH99,Mark16} for any state $\sigma$ with dimension $d$ is defined as 
\begin{equation} \label{DepChDis}
\Lambda^{\text{dep}} \left( \sigma \right) = p \sigma + (1-p) \frac{\mathbb{I}}{d}.
\end{equation}
A depolarizing channel $\Lambda^{\text{dep}}$ is equivalent to noisy singlets denoted by $\rho_{p}$ can be written as
\begin{equation} \label{Lam_dep_rho_NS}
\Lambda^{\text{dep}} \equiv \rho_{p}.
\end{equation}

\item Any channel $\Lambda$ with subjected to a twirling process denoted by $\mathcal{T}$ becomes a depolarizing channel as
\begin{equation} \label{T_Lam_Lam_dep}
\mathcal{T} \left(\Lambda \right)  = \Lambda^{\text{dep}}.
\end{equation}

\item For any channel $\Lambda$, let us denote the channel fidelity by $f(\Lambda)$ and the entanglement fidelity by $F(\Lambda)$. Now using the equivalence between depolarizing channel and noisy singlets as shown in Eq.~\eqref{Lam_dep_rho_NS}, we can write the relation between teleportation fidelity and singlet fraction in Eq.~\eqref{f_F_par} for any depolarizing channel $\Lambda^{\text{dep}}$ as
\begin{equation} \label{f_F_Lam_dep}
f(\Lambda^{\text{dep}})= \frac{F(\Lambda^{\text{dep}}) d +1}{d+1}.
\end{equation} 

\item Further, it can be proved that the teleportation fidelity $f$ and the singlet fraction $F$ are invariant of a twirling process $\mathcal{T}$. Thus for any channel $\Lambda$, we can write Eq.~\eqref{f_F_Lam_dep} as
\begin{equation}\label{f_F_Lam}
f(\Lambda)= \frac{F(\Lambda) d +1}{d+1}.
\end{equation}

\item Now from the isomorphism between channel and state in Eq.~\eqref{Lab_rho}, we can write Eq.~\eqref{f_F_Lam} as
\begin{equation}
f(\rho)= \frac{F(\rho) d +1}{d+1}.
\end{equation}
\end{enumerate} 
 Thus, form the above statements we can conclude that,  as the relation between the teleportation fidelity $(f)$ and  the singlet fraction $(F)$ in Eq.~\eqref{f_F_par} are true for depolarising channel, is also true for any channel. Finally, from the channel-state isomorphism, one can conclude that the relation in Eq.~\eqref{f_F_par} is also true for any state.

\section{Details of device-independent quantum private query protocols.} \label{DIQPQapp}
Quantum private query (QPQ)~\cite{Giovannetti08} is a cryptographic protocol that deals with the communication between a database owner and its clients. Registered clients can query about required entries in the database and the server returns the appropriate answer to that query. The security in this protocol depends upon two things:
\begin{enumerate}
\item The server should not reveal any extra information to the clients other than their queries.
\item The server should not gain any extra information about the queries of the clients. 
\end{enumerate}
Conventionally, it is assumed Bob is the database owner or the server and Alice is the client. Like other cryptographic protocols, the communication will start with the key establishment part. However there is a major difference with normal quantum key distribution protocols where the whole key is shared between Alice and Bob. Here, the key is distributed between Alice and Bob such that
\begin{enumerate}
\item Bob, the database owner knows the whole key.
\item Alice, the client knows only a part of the key or the value of some specific positions of the key.
\item The database owner has no information about the positions of the key known to the clients.	
\end{enumerate}
As a result, unlike normal quantum key distribution protocols, there is no need for any external advisories other than Alice and Bob such as eve. Here, Alice and Bob both may work like adversaries to each other. The motivation of Alice will be to gain more information about the database whereas Bob will try to know the position of the bits of the key known to Alice and hence gain information about her queries. 
\subsubsection{Quantum private query protocol}
Here, we will discuss the QPQ protocol as proposed in~\cite{Yang14} which is based on B92 quantum key distribution~\cite{B92} scheme. There are two phases 
\begin{enumerate}
\item \textit{Key generation:} In this phase, a secure key is established between Bob and Alice. This can be done in two ways:
\begin{enumerate}
\item The source shares an entangled pair of particles with Bob and Alice in the specific form
\begin{equation} \label{QPQstate}
\ket{\psi}_{BA}=\frac{1}{\sqrt{2}}\left( \ket{0}_{B}\ket{\phi_{0}}_{A} + \ket{1}_{B}\ket{\phi_{1}}_{A}\right)
\end{equation}
where
\begin{eqnarray*}
\ket{\phi_{0}}_{A} = \text{cos} \left( \frac{\theta}{2}\right) \ket{0} + \text{sin} \left( \frac{\theta}{2}\right) \ket{1}, \\
\ket{\phi_{1}}_{A} = \text{cos} \left( \frac{\theta}{2}\right) \ket{0} - \text{sin}\left( \frac{\theta}{2}\right) \ket{1},
\end{eqnarray*}
and $0 < \theta < \frac{\pi}{2}$.

\item Now, after sharing Bob measures his qubits in $\left\lbrace \ket{0}_{B}, \ket{1}_{B} \right\rbrace $ basis and Alice measures her qubits in either $\left\lbrace \ket{\phi_{0}}_{A}, \ket{\phi_{0}^{\perp}}_{A} \right\rbrace $ basis or $\left\lbrace \ket{\phi_{1}}_{A}, \ket{\phi_{1}^{\perp}}_{A} \right\rbrace $ basis.
\end{enumerate}
By simple calculations~\cite{Yang14}, it can be concluded if Alice's measurement output is $\ket{\phi_{0}^{\perp}}$ or $\ket{\phi_{1}^{\perp}}$, then Bob's measurement output must be $1$ or $0$ respectively. After classical post-processing, in this way, a key can be established between Alice and Bob in such a way that Alice can get only one or more bits of information of the whole key. In contrast, Bob has knowledge of the whole key but Bob has no information about the bits or bits of the key known to Alice.
 
\item \textit{Private query:} After the key is established let $K$, then let us assume Alice knows the $j$ th bit of key and she makes a query about the $i$th element of the database. Then she calculated an integer $s = (j - i)$. Alice sends $s$ to Bob. He then shifts the key $K$ by $s$ amount and generates a new key, say $\bar{K}$. Using this new key $\bar{K}$, Bob encrypts his database using a one-time pad. Bob then transmits the whole database to Alice. Alice can easily get the $j$th bit by decrypting the database.
\end{enumerate}
\subsubsection{Device  independent tests}
The security of the above protocol lies in the fact that the source shares the specific state with the Bob as given in Eq.~\eqref{QPQstate} for a specific value of $\theta$. It can be shown that if the state is not exactly the same, i.e., for some value $\theta + \epsilon$ where $\epsilon \neq 0$, Alice can always generate more information~\cite{DIQPQ}. To mitigate this problem, Bob has to remove the trust in the source and should test the correctness of the state given to him at his end. There are several tests to check the correctness of the given state. The test is chosen which gives the maximum success probability. 
\begin{enumerate}
\item \textit{Local Clauser-Horne-Shimony-Holt (CHSH) test}: Bob has to perform this local CHSH test on some of the randomly chosen $n$  pairs. The main steps are
\begin{enumerate}
\item Bob chooses two random bit strings $x_i, y_i \in \lbrace0, 1\rbrace$ where $i \in \lbrace1, 2, \ldots, n\rbrace$.
\item If $x_i = 0$,  Bob measures the first particle in $\lbrace \ket{0}, \ket{1}\rbrace$ basis, else in $\lbrace \ket{+}, \ket{-}\rbrace$ basis.
\item Similarly, if $y_i=0$, the the Bob measures the first particle in $\lbrace \ket{\psi_1}, \ket{\psi^{\perp}_1}\rbrace$ basis, else in $\lbrace \ket{\psi_2}, \ket{\psi^{\perp}_2}\rbrace$ basis where
\begin{eqnarray*}
\ket{\psi_1} = \text{cos} \left( \frac{\psi_1}{2}\right) \ket{0} + \text{sin} \left( \frac{\psi_1}{2}\right) \ket{1} , \\ \ket{\psi_2} = \text{cos} \left( \frac{\psi_2}{2}\right) \ket{0} + \text{sin} \left( \frac{\psi_2}{2}\right) \ket{1}.
\end{eqnarray*}
\item The measurement result is stored in another bit stings $a_i, b_i \in \lbrace0, 1\rbrace$ such that if the measurement result of fist particle is $\ket{0}$ or $\ket{+}$, then $a_i=0$, else $a_i=1$. 
\item Similarly, if the measurement result of the second particle is $\ket{\psi_1}$ or $\ket{\psi_2}$, then $a_i=0$, else $a_i=1$.
 \item The test is called successful if $a_i \otimes b_i = x_i \wedge y_i$.
\end{enumerate} 
The success probability of this test is 
\begin{equation}
P_{s} =\frac{1}{8} \left(\text{sin}\theta \left(\text{sin} \psi_1 + \text{sin} \psi_2 \right) + \text{cos}\psi_1 - \text{cos}\psi_2\right)+\frac{1}{2} 
\end{equation}
 which is dependent on the values of $\theta$, $\psi_1$, and $\psi_2$. The maximum value of $P_{s}$ is 0.85.
 \item \textit{Quantum pseudo-telepathy test}: Using this test the success probability can reach upto unity. The steps of this test are given below
 \begin{enumerate}
 \item With the help of an ancilla qubit $X$, the state in Eq.~\eqref{QPQstate} can be transformed into the following state
 \begin{widetext}
 \begin{equation} \label{pseudo_state}
 \ket{\psi}_{BAX} = \frac{1}{\sqrt{2}} \left( \text{cos}\frac{\theta}{2} \ket{000}_{BAX}  + \text{sin}\frac{\theta}{2} \ket{010}_{BAX} + \text{cos}\frac{\theta}{2} \ket{111}_{BAX} - \text{sin}\frac{\theta}{2} \ket{100}_{BAX} \right). 
 \end{equation}
 \end{widetext}
 \item Bob will now randomly choose a three-bit number, ensuring that the string has an even number of 1's. Subsequently, depending on the input, he conducts specific measurements on the state described in Eq.~\eqref{pseudo_state}. Using the resulting outputs, he must generate an output containing an even number of 1's only when the number of 1's in the input is divisible by 4.
 \end{enumerate}
Now using some specific measurements as given in~\cite{Jyoti18} it can be shown that the success probability is $P_{s}=\frac{1}{4} \left(3 + \text{cos} \theta \right) $. This value goes to 1 asymptomatically as $0 < \theta < \frac{\pi}{2}$.
\end{enumerate}

\end{document}